\documentclass[journal]{IEEEtran}

\usepackage{xcolor}
\newcommand\hl[1]{%
	\bgroup
	\hskip0pt\color{red!80!black}#1%
	\egroup
}
\newcommand\hsig[0]{%
	\bgroup
	\hspace{.3em}
	\egroup
}

\usepackage{multicol,multirow,ragged2e,graphicx,color}
\usepackage{color,caption}
\usepackage{subcaption}

\begin{document}
\title{LNDb: A Lung Nodule Database on Computed Tomography}

\author{Jo\~ao~Pedrosa,
        Guilherme~Aresta,
        Carlos~Ferreira,
        M\'arcio~Rodrigues,
        Patr\'icia~Leit\~ao,
        Andr\'e~Silva~Carvalho,
        Jo\~ao~Rebelo,
        Eduardo~Negr\~ao,
        Isabel~Ramos,
        Ant\'onio~Cunha,
        and~Aur\'elio~Campilho,
\thanks{Manuscript received \today; revised \today.}
\thanks{J. Pedrosa, G. Aresta, C. Ferreira, A. Cunha and A. Campilho are with the Institute for Systems and Computer Engineering, Technology and Science (INESC TEC), Porto, Portugal, e-mail: joao.m.pedrosa@inesctec.pt.}
\thanks{G. Aresta and A. Campilho are also with the Faculty of Engineering of the University of Porto (FEUP), Porto, Portugal}
\thanks{A. Cunha is also with the University of Tr\'{a}s-os-Montes e Alto Douro (UTAD), Vila Real, Portugal}
\thanks{M. Rodrigues, P. Leit\~ao, A. S. Carvalho, J. Rebelo, E. Negr\~ao and I. Ramos are with the Department of Radiology, Centro Hospitalar e Universit\'{a}rio de S. Jo\~ao, Porto, Portugal.}
\thanks{I. Ramos is also with the Faculty of Medicine of the University of Porto (FMUP), Porto, Portugal}}

\markboth{Arxiv, \today}%
{Pedrosa \MakeLowercase{\textit{et al.}}: TITLE}

\maketitle

\begin{abstract}
Lung cancer is the deadliest type of cancer worldwide and late detection is the major factor for the low survival rate of patients. Low dose computed tomography has been suggested as a potential screening tool but manual screening is costly, time-consuming and prone to variability. This has fuelled the development of automatic methods for the detection, segmentation and characterisation of pulmonary nodules but its application to clinical routine is challenging. In this study, a new database for the development and testing of pulmonary nodule computer-aided strategies is presented which intends to complement current databases by giving additional focus to radiologist variability and local clinical reality. State-of-the-art nodule detection, segmentation and characterization methods are tested and compared to manual annotations as well as collaborative strategies combining multiple radiologists and radiologists and computer-aided systems. It is shown that state-of-the-art methodologies can determine a patient's follow-up recommendation as accurately as a radiologist, though the nodule detection method used shows decreased performance in this database.
\end{abstract}

\begin{IEEEkeywords}
lung cancer, low dose computed tomography, pulmonary nodules, computer-aided diagnosis, deep learning.
\end{IEEEkeywords}

\IEEEpeerreviewmaketitle

\section{Introduction}
\IEEEPARstart{L}{ung} cancer is the deadliest type of cancer worldwide for both men and women \cite{siegel2019cancer}. Though changes in the smoking patterns in the general population have been largely responsible for decreasing trends in incidence and mortality rates in recent decades, lung cancer is still responsible for over double the cancer deaths of colorectal cancer, the second deadliest cancer type, and is projected to remain the deadliest type of cancer in the near future. Progress in increasing lung cancer survival rate has also been notoriously slow in contrast to other cancer types, mainly due to late diagnosis of the disease. 
Low-dose computed tomography (CT) has long been suggested as a potential early screening tool and a 20\% reduction in lung cancer mortality has been demonstrated for lung cancer risk groups \cite{aberle2011national}. Nevertheless, translation of these screening programs to the general population has been challenging due to equipment and personnel costs and the complexity of the task. Namely, lung nodules present a large range of shapes and characteristics and thus the identification and characterization of these abnormalities is not trivial and prone to high interobserver variability. Computer-aided diagnosis (CAD) systems can thus facilitate the adoption and generalization of screening programs by reducing the burden on the clinicians and providing a second-opinion.

Extensive research has been conducted on the development of CAD systems for lung cancer screening focusing on the different tasks essential for efficient screening - pulmonary nodule detection and segmentation followed by nodule characterization and classification of malignancy. Recently, deep learning based methods have shown especially promising results for nodule detection \cite{ding2017accurate,aresta2018towards,wang2017central,wu2018joint}, segmentation \cite{wang2017central,wu2018joint,aresta2018towards,yaguchi20193d} and characterization \cite{wu2018joint,ferreira2018convolutional,shen2019interpretable,ferreira2019wide}. In fact, most of the best performing methods on the LUNA16 nodule detection challenge use deep learning \cite{setio2017validation} and the same trend was observed for detection and malignancy classification on Kaggle's Data Science Bowl 2017 challenge \cite{kaggle2017}.

Given the dependence of deep learning methods on large datasets with robust ground truth, the publication of annotated datasets has been a hugely important contribution for the community. Perhaps the most widely known public database is the LIDC-IDRI \cite{armato2011lung}, which contains 1018 CT scans, each annotated by four radiologists. Annotations comprise nodule segmentation and subjective characterization \cite{mcnitt2007lung}, making this an extremely useful database for the development of CAD approaches in lung cancer screening. The NLST database is also widely recognised and contains CTs from 26.722 patients, though nodule segmentation and characterization are not available and nodule position is limited to the slice where a nodule was found \cite{aberle2011national}.

In spite of the promising results in literature, adoption of CAD systems as part of a broader screening in the clinic is not straightforward. First, the fact that CAD systems are not designed as an integrated part of the clinical routine makes them difficult to adopt. In fact, if not well integrated into the normal routine, they come to represent an extra step in the pipeline, increasing the burden on clinicians. Secondly, in spite of the large quantity and variety of data in a dataset like LIDC-IDRI, translation to a local reality can present challenges such as different acquisition settings, population demographics, pathologies or others, which can be detrimental to the performance of deep learning methods, making a local validation of any CAD system an essential step.

To tackle these issues, the Lung Nodule Database (LNDb) was developed as an external dataset complimentary to LIDC-IDRI. The publication of this database will give continuity to LIDC-IDRI and allow the community to perform an external and comparable validation of proposed CAD systems. Furthermore, the fact that eyetracking was used during manual annotation of the images (cf. Section \ref{ssec:annotmethods}) allows for the development of collaborative strategies for CAD, ensuring that CAD systems are designed as allies of radiologists rather than as competition.

\section{Methodology}
\subsection{Patient Selection and Data Acquisition}
The LNDb contains 294 CT scans collected retrospectively at the Centro Hospitalar e Universit\'{a}rio de S\~ao Jo\~ao (CHUSJ) in Porto, Portugal between 2016 and 2018. All data was acquired under approval from the CHUSJ Ethical Commitee and was anonymised prior to any analysis to remove personal information except for patient birth year and gender. No scan was acquired specifically for LNDb.

To ensure that the database is relevant, inclusion criteria based on the LIDC-IDRI criteria were used \cite{armato2011lung}. All patients above the age of 18 were included, except if a prior history of cancer was known. CT scans were collected patientwise to avoid repeated patients. CT scans where intravenous contrast had been used and those with a slice thickness greater than 1mm were excluded. One radiologist then performed a reading of the CT to look for other lung pathologies, noise, motion or other artifacts, in which case the CT would be excluded. If during the first reading more than six nodules or one nodule larger than 30mm in-slice diameter were found the CT would also be excluded. Finding more than six nodules during image annotation was not a reason for exclusion.

Table \ref{tb:acqsettings} shows the acquisition parameters for the CT scans in LNDb. Among the 294 patients scanned, 164 (55.8\%) were male. The median age was 66 and the minimum and maximum ages were 19 and 98, respectively.

\begin{table}
	\centering
	\begin{tabular}{lc}
		\hline
		\vspace{-.2cm}&\\
		Scanner Model&\\
		\hspace{.3cm} Siemens Sensation Cardiac 64\textsuperscript{a} & 107 (36.4)\\
		\hspace{.3cm} Siemens Somatom Definition Flash\textsuperscript{a} & 59 (19.5)\\
		\hspace{.3cm} Siemens Somatom go.Up\textsuperscript{a} & 137 (45.2)\\
		Tube Peak Potential (kV)\textsuperscript{b} & 120[100;140]\\
		Average Tube Current (mA)\textsuperscript{c} & 161.9$\pm$128.4\\
		Convolution kernel&\\
		\hspace{.3cm} Standard\textsuperscript{a} & 4 (1.4)\\
		\hspace{.3cm} Sharp\textsuperscript{a} & 160 (54.6)\\
		\hspace{.3cm} Very sharp\textsuperscript{a} & 126 (43.0)\\
		\hspace{.3cm} Extremely sharp\textsuperscript{a} & 3 (1.0)\\
		In-plane pixelsize (mm)\textsuperscript{c} & 0.63$\pm$0.09\\
		Slice thickness (mm)\textsuperscript{b} & 1.0[0.5;1.0]\\
		Number of image slices\textsuperscript{b} & 318.5[251;631]\\
		\hline
		\vspace{-.2cm}&\\	
	\end{tabular}
	\\
	\centering
	\textsuperscript{a}Data are count (\%);
	\textsuperscript{b}Data are median[minimum;maximum];\\
	\textsuperscript{c}Data are mean$\pm$standard deviation.
	\caption{CT scan acquisition settings.}
	\label{tb:acqsettings}
\end{table}

\subsection{Manual Annotation Process}
\label{ssec:annotmethods}
\begin{figure*}
	\centering
	\begin{subfigure}[c]{0.32\textwidth}
		\includegraphics[width=\textwidth,trim={.8cm 1cm .8cm 3cm},clip]{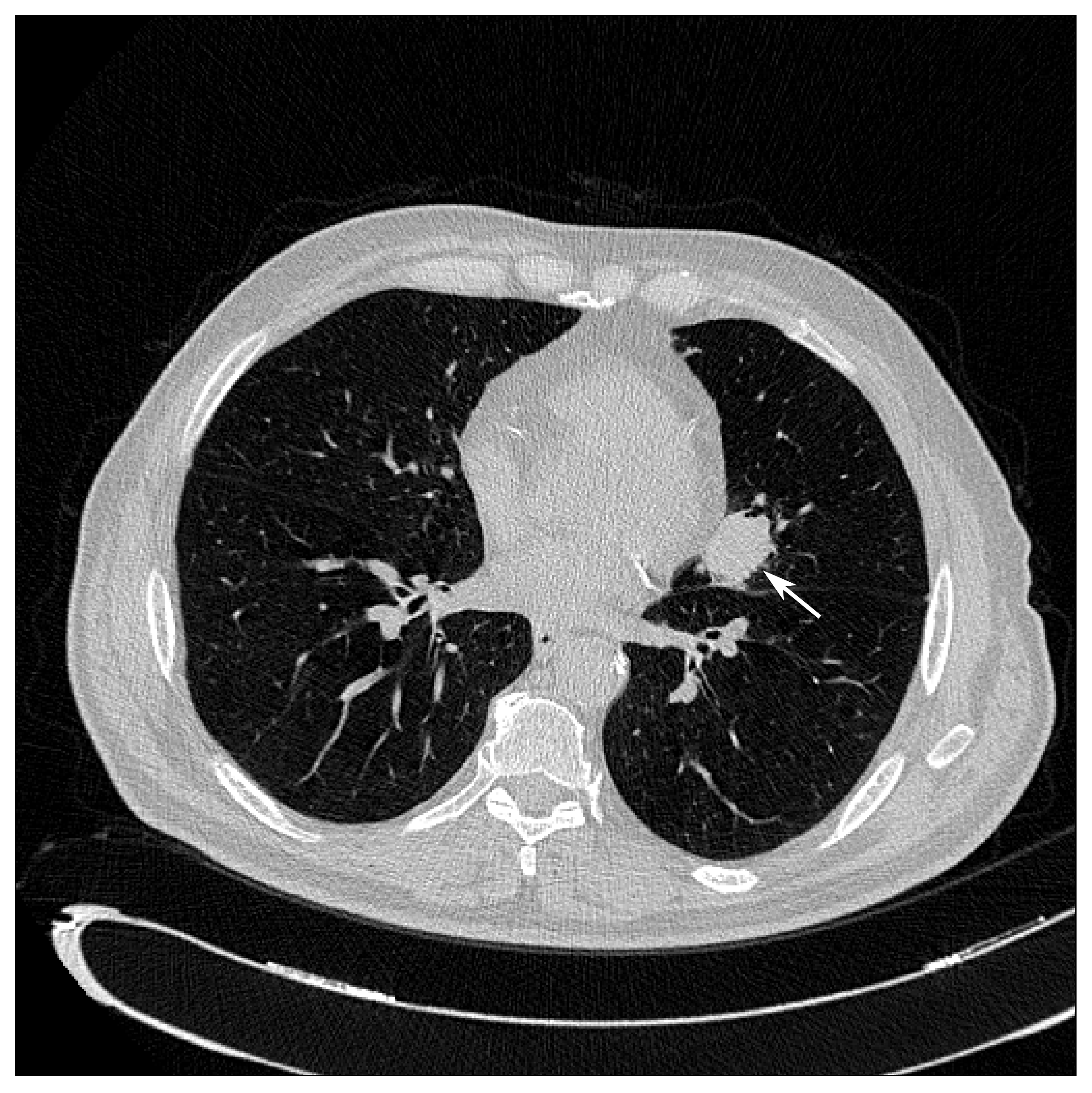}
		\vspace{-.6cm}
		\caption{}
		\vspace{-.2cm}
	\end{subfigure}
	\begin{subfigure}[c]{0.32\textwidth}
		\includegraphics[width=\textwidth,trim={.8cm 0.8cm .8cm 0.8cm},clip]{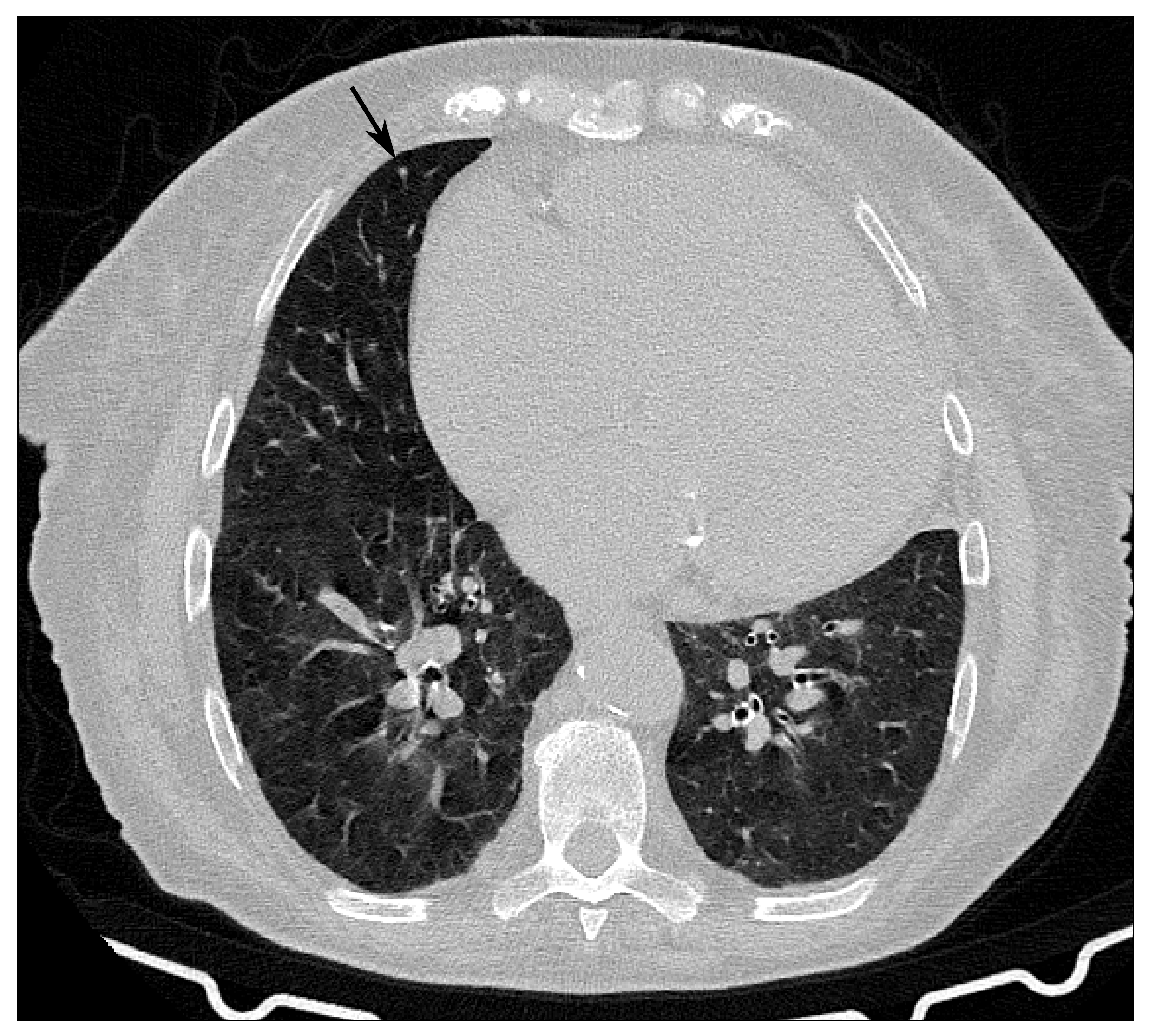}
		\caption{}
		\vspace{-.3cm}
	\end{subfigure}
	\begin{subfigure}[c]{0.32\textwidth}
		\includegraphics[width=\textwidth,trim={.8cm 1cm .8cm 3cm},clip]{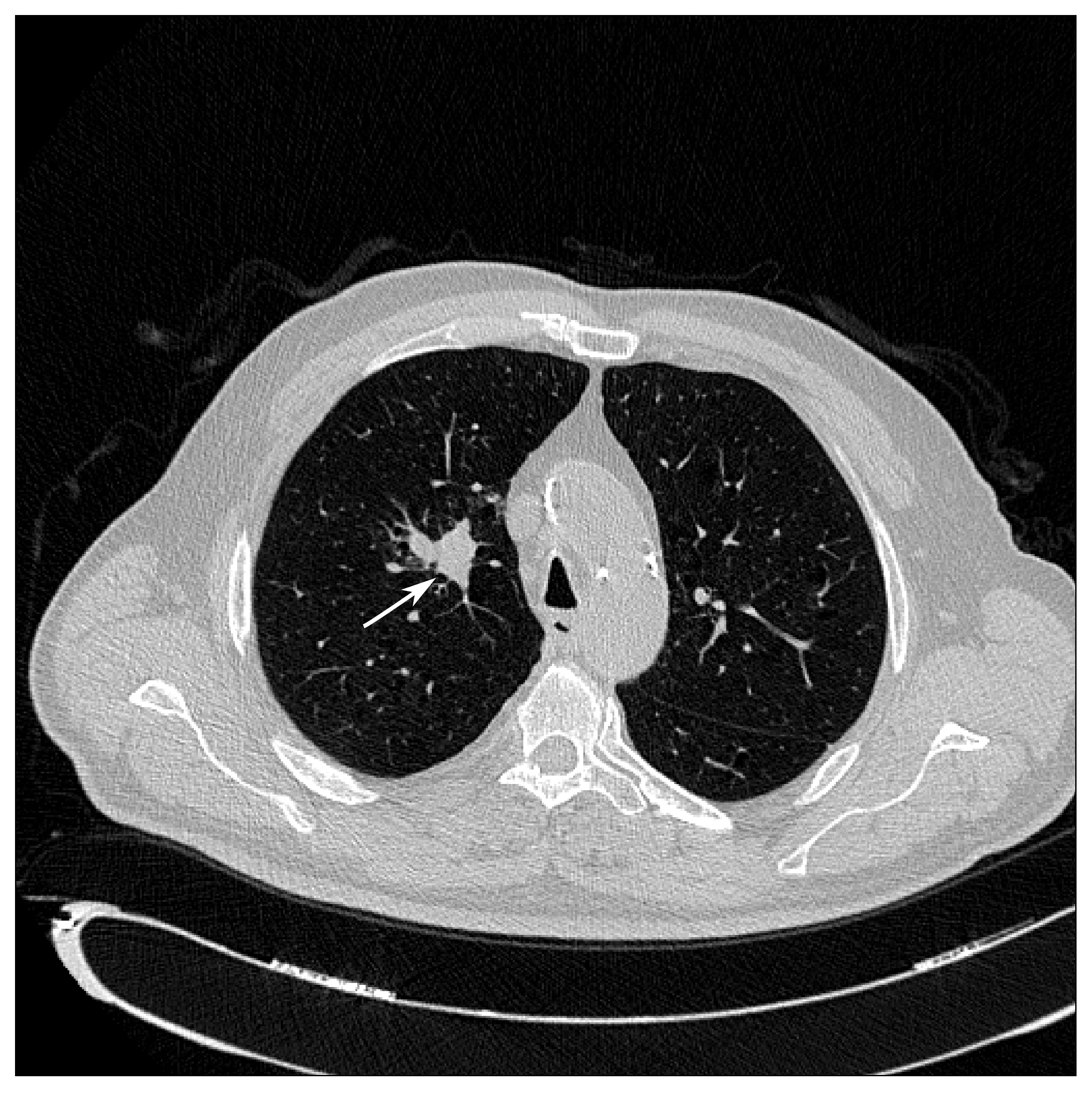}
		\caption{}
		\vspace{-.3cm}
	\end{subfigure}
	
	\caption{\small Examples of annotated lesions. (a) Nodule $\geq$3mm annotated by 3 radiologists; (b) Nodule $<$3mm annotated by 2 radiologists; (c) Non-nodule annotated by 2 radiologists.}
	\label{fig:LNDbAnnot}
\end{figure*}

Each CT scan was read by at least one radiologist at CHUSJ to identify pulmonary nodules and other suspicious lesions. A total of 5 radiologists with at least 4 years of experience reading up to 30 CTs per week, hereinafter referred to as R1 to R5, participated in the annotation process. Annotations were performed in a single blinded fashion, i.e. a radiologist would read the scan once and no consensus or review between the radiologists was performed. Each scan was read by at least one radiologist. The instructions for manual annotation were adapted from LIDC-IDRI\cite{armato2011lung}. Each radiologist would read a scan and identify the following lesions: i) nodule $\geq$3mm: any lesion considered to be a nodule by the radiologist with greatest in-plane dimension larger or equal to 3mm; ii) nodule $<$3mm: any lesion considered to be a nodule by the radiologist with greatest in-plane dimension smaller than 3mm; iii) non-nodule: any pulmonary lesion considered not to be a nodule by the radiologist, but that contains features which could make it identifiable as a nodule. Figure \ref{fig:LNDbAnnot} show examples of annotated lesions in LNDb.

Nodules $\geq$3mm were segmented and subjectively characterized according to LIDC-IDRI (ratings on subtlety, internal structure, calcification, sphericity, margin, lobulation, spiculation, texture and likelihood of malignancy). For a complete description of these characteristics the reader is referred to McNitt-Gray et al. \cite{mcnitt2007lung}. For nodules $<$3mm the nodule centroid was marked and subjective assessment of the nodule's characteristics was performed. For non-nodules, only the lesion centroid was marked.

Manual annotation was performed in an in-house developed graphical interface \cite{pedrosa2019medicon}. It allows for image orientation and magnification as well as selection of different display windows. Maximum intensity projection was used only for a portion of the scans as it was not available from the start of the project. Lesion segmentation and subjective assessment were performed on the axial slice but radiologists had access to the coronal and sagittal slices as well. Lesion segmentation was performed manually by using a brush to color the nodule.

Eyetracking was performed during manual annotation to record the radiologists' gaze \cite{machado2018radiologists}. The screen coordinates of the radiologists' gaze were recorded at a frequency of 90Hz together with the display settings of the graphical interface. This allows conversion of the screen coordinates to CT image coordinates taking into account zoom and pan setting, thus allowing to compute which region of the CT the radiologist was looking at throughout the annotation process. Considering a 5$^{\circ}$ visual angle \cite{millodot2014dictionary}, a 3D attention map of the radiologists' annotation process can be reconstructed. This 3D attention map can then be used to compute the amount of time spent at each image location.

\subsection{Computer-Aided Annotation}
\label{ssec:CADmethods}
For comparison to manual annotation, previously developed methods for computer-aided detection, segmentation and characterisation of nodules were used. All methods were developed and trained exclusively on nodules $\geq$3mm from LIDC-IDRI.

The nodule detection approach is based on the YOLOv3 architecture \cite{redmon2018yolov3}. In brief, a model pre-trained in natural images is fine-tuned to detect lung nodules by minimizing a loss function that takes into account the width, height, and centroid of the prediction in relation to the ground truth. To account for 3D information, the algorithm is trained with 3-channel images composed of the axial slice containing the nodule's center of mass and two equidistant adjacent slices \cite{aresta2018towards}. Predictions are performed for every axial slice and the candidates are merged if their bounding boxes overlap.

After candidate detection, a dedicated network for false positive (FP) reduction is used. The network is composed of blocks of 3$\!\times\!$3$\!\times\!$3 convolutions with batch normalization and rectifier linear unit activations. The input size is a cube of size 64$\!\times\!$64$\!\times\!$64 voxels centered on the candidate centroid. The binary non-nodule/nodule classification is considered as a multiple-instance learning problem so that the probability is inferred by max-pooling on a 8$\!\times\!$8$\!\times\!$8$\!\times\!$1 feature map. The training dataset is composed of all nodules used for training the detection network as well as the highest scored FPs from each scan in a 1:5 ratio.

The segmentation network is iW-Net\cite{aresta2019iw}, a model that allows for both automatic and semi-automatic segmentation. The model is composed of two sequential auto-encoders based on the 3D U-Net \cite{cciccek20163d} that receives as input a 64$\!\times\!$64$\!\times\!$64 voxel candidate. The first block predicts an initial segmentation, which can then be refined by the second block that assesses the nodule image, the initial segmentation and a weight map resulting from two manual clicks near the nodule boundaries. In this study, only the first block of iW-Net is used.

For nodule characterisation, only texture was used. Three orthogonal planes of 64$\!\times\!$64 pixels centered on the candidate center of mass, are given as input to a convolutional neuronal network. The features extracted from each of the three planes are then concatenated so that there is a common output with 3 classes: ground glass opacity (GGO), part solid and solid \cite{ferreira2018convolutional}.

\section{Experiments}
\subsection{Observer Variability}
To assess interobserver variability, the annotations of multiple radiologists on matching CTs were compared. Two annotations by different radiologists were considered to correspond to the same lesions, i.e. a unique finding, if the Euclidean distance between their centroids was smaller or equal to the maximum equivalent diameter of the two nodules. For nodules of equivalent diameter smaller than 3mm, an equivalent diameter of 3mm was considered.

Nodule detection agreement was computed as the percentage of cases in agreement over all findings reported as a nodule by at least one of the radiologists being considered:
\begin{equation}
A_d = \frac{n_{N,N}}{n_{N,N}+n_{N,N\!N}+n_{N\!N,N}},
\end{equation}
where $n_{X,Y}$ is the number of findings reported as class $X$ by radiologist 1 and class $Y$ by radiologist 2. $N$ and $N\!N$ are the ``nodule" and ``non-nodule/not reported" classes, respectively.

Nodule segmentation agreement was evaluated through Jaccard score \cite{levandowsky1971distance}, Hausdorff distance (HD) \cite{huttenlocher1993comparing} and mean average distance (MAD) computed as
\begin{equation}
M\!AD = \frac{1}{2}(d(S_1,S_2)+d(S_2,S_1)),
\end{equation}
where $d(S_1,S_2)$ is the mean of the distance between each surface voxel in segmentation $S_1$ and the closest surface voxel in segmentation in $S_2$; $d(S_2,S_1)$ is computed in the same way.

Nodule characterization agreement was evaluated for each characteristic using Fleiss-Cohen weighted Cohen's kappa \cite{spitzer1967quantification}
\begin{equation}
\kappa_w=\frac{\sum_{i}^{k}\sum_{j}^{k}w_{ij}p_{ij}-\sum_{i}^{k}\sum_{j}^{k}w_{ij}p_{i*}p_{*j}}{1-\sum_{i}^{k}\sum_{j}^{k}w_{ij}p_{i*}p_{*j}}
\end{equation}
where $p_{ij}$ is the proportion of cases rated by observer 1 as class $i$ and by observer 2 as class $j$. $*$ is a wildcard so that $p_{*j}$ is the proportion of cases rated by observer 2 as class $j$. $w_{ij}$ is the weight for class combination $ij$ according to
\begin{equation}
w_{ij}=\frac{(C_i-C_j)^2}{(C_1-C_k)^2}
\end{equation}
for a rating consisting of $k$ classes ($C_1$,$C_2$,...,$C_k$). Note that for internal structure and calcification, the non-weighted Cohen's kappa $\kappa$ is reported given the non-ordinal nature of these features. Given that for LIDC-IDRI the radiologists' identity is unknown and Cohen's kappa cannot be computed, in-class agreement ($A_c$) is reported for comparison, where as the proportion of cases rated the same class by both observers.

As a measure for scanwise agreement, the Fleischner society pulmonary nodule guidelines \cite{macmahon2017guidelines} were used to obtain follow-up recomendations for each CT scan according to the annotations of each radiologist. The Fleischner guidelines are widely used for patient management in the case of nodule findings and take into account the number of nodules (single or multiple), their volume ($<\!\!100mm^3$, $100\!\!-\!\!250mm^3$ and $\geq\!\!250mm^3$) and texture (solid, part solid and GGO nodules). Nodule volume was computed from the segmentation and nodules $<$3mm were considered to belong to the first class ($<\!\!100mm^3$). Nodule texture was recast from the five classes in the LNDb annotation (1-GGO, 2-intermediate, 3-part solid, 4-intermediate, 5-solid) into the three classes of the Fleischner guidelines by considering GGO as 1-2, part solid as 3 and solid as 4-5. The Fleischner follow-up guidelines were then divided into 4 classes of escalating risk: 0) No routine follow-up required or optional CT at 12 months according to patient risk; 1) CT at 6-12 months required; 2) CT at 3-6 months required; 3) CT, PET/CT or tissue sampling at 3 months required.

This 0-3 score, hereinafter referred to as Fleischner score, was then used to compare the follow-up recommendation as assessed by each radiologist using $\kappa_w$. The agreement per nodule in the volume and texture classes was also assessed using $\kappa_w$ and $A_c$.


\subsection{Computer-Aided Annotation}
\label{ssec:CADexp}
In order to assess the performance of state-of-the-art CAD methods in relation to the radiologists' manual annotations, automatic nodule detection, segmentation and characterization was performed in all CTs.

Nodule detection performance was evaluated in terms of sensitivity and number of FPs per scan as function of the FP reduction threshold. Given that not all findings were annotated by all radiologists, performance was assessed in relation to the radiologists' agreement level, considering findings marked as a nodule by at least one or at least two radiologists. Segmentation performance was evaluated in terms of MAD, HD and Jaccard and characterization performance in terms of $A_c$ and $\kappa_w$. Finally, scanwise performance for the full CAD pipeline in terms of Fleischner score was evaluated using $\kappa_w$. Similarly to the nodule detection performance evaluation, the agreement level was taken into consideration by computing the radiologists' Fleischner score taking into consideration findings marked by at least one or by at least two radiologists.

\subsection{Collaborative Annotation Strategies}
In order to assess the feasibility of collaborative CAD systems, a 2nd opinion experiment was conducted in 23 randomly selected cases. After manual annotation by two radiologists (R4 and R5), each radiologist received suggestions for revision from the other radiologist and the CAD system in terms of nodule detection, segmentation and texture characterization. Suggestions from the other radiologist and CAD were blinded so that each radiologist would not know the source of each suggestion and thus avoid bias in the decision process.

For nodule detection comparison, each radiologist received as suggestions for revision all findings marked as nodules by the other radiologist or CAD if the radiologist had marked it as non-nodule or had not reported it. For nodule segmentation comparison, the LIDC-IDRI interobserver nodule segmentation variability was used to determine nodules in disagreement. As such, when comparing two nodule segmentations, if they belonged to the same Fleischner volume class ($<\!\!100mm^3$, $100\!\!-\!\!250mm^3$ and $\geq\!\!250mm^3$) and had a HD outside the LIDC-IDRI HD variability by 2 standard deviations or if they belonged to different Fleischner volume classes and had a HD outside the LIDC-IDRI HD variability by 1 standard deviation they would be presented for revision to the radiologist. For nodule characterization, a nodule would be presented for revision if the annotated nodule textures did not belong to the same Fleischner texture class (1-2, 3 and 4-5).

After revision of nodule detection, segmentation and texture, the revised annotations by each radiologist were compared and cases in disagreement were revised by both radiologists together to obtain a consensus ground truth. In the particular case of segmentation, the ground truth in nodules which were not revised in the consensus phase was considered to be the average volume of the revised segmentation of the two radiologists.

For each revised annotation by a radiologist, the contributions from the other radiologist and the CAD system were then disentangled to allow for a separate assessment of the different annotation strategies possible: i) single radiologist or CAD annotation; ii) first radiologist annotation followed by revision of second radiologist findings; iii) single radiologist annotation followed by revision of CAD findings. Note that for strategy iii) the number of CAD findings received by the radiologist can be regulated by adjusting the FP threshold used during detection. Furthermore, for strategies ii) and iii), the number of findings received by the radiologist can be regulated by removing findings according to the time spent on that finding's region during the initial image annotation. The amount of time spent in the region around each finding during manual annotation was computed from the eyetracking map and findings for which the time spent in the region was superior to a predetermined attention threshold were excluded.

Detection, segmentation and characterization performance were then evaluated in regard to the consensus annotations by R4 and R5. As in Section \ref{ssec:CADexp}, detection performance was evaluated in terms of sensitivity and FPs per scan. To assess the burden for the clinicians associated with each strategy, average time expenditure per scan was assessed for each strategy through the eyetracking map. Segmentation and texture characterization performance were evaluated in terms of $A_c$ and $\kappa_w$ in the Fleischner volume and texture classes and scanwise performance was assessed in terms of $\kappa_w$.

Finally, CAD candidates (at an FP threshold of 0.5) identified as FPs by the radiologists were revised by R4 to identify the anatomical features which were most responsible for FPs.

\subsection{Statistical Analysis}
For comparison between different observers and databases, unpaired t-tests were used taking into account significance at p$<$0.05 and p$<$0.01.

\section{Results}
\subsection{LNDb Description}
All 294 CTs of LNDb were annotated by at least one radiologist (90 were annotated by 3 radiologists, 145 by 2 radiologists and 59 by a single radiologist). R1 to R5 annotated respectively 125, 90, 81, 162 and 161 CTs. Eyetracking data was collected for a total of 312 CT readings. The database comprises 1897 annotations by the 5 radiologists, corresponding to 1429 unique findings.

\begin{figure*}
	\centering
	\begin{subfigure}[c]{0.36\textwidth}
		\includegraphics[width=\textwidth,trim={0.2cm 0.3cm 0.3cm 0.2cm},clip]{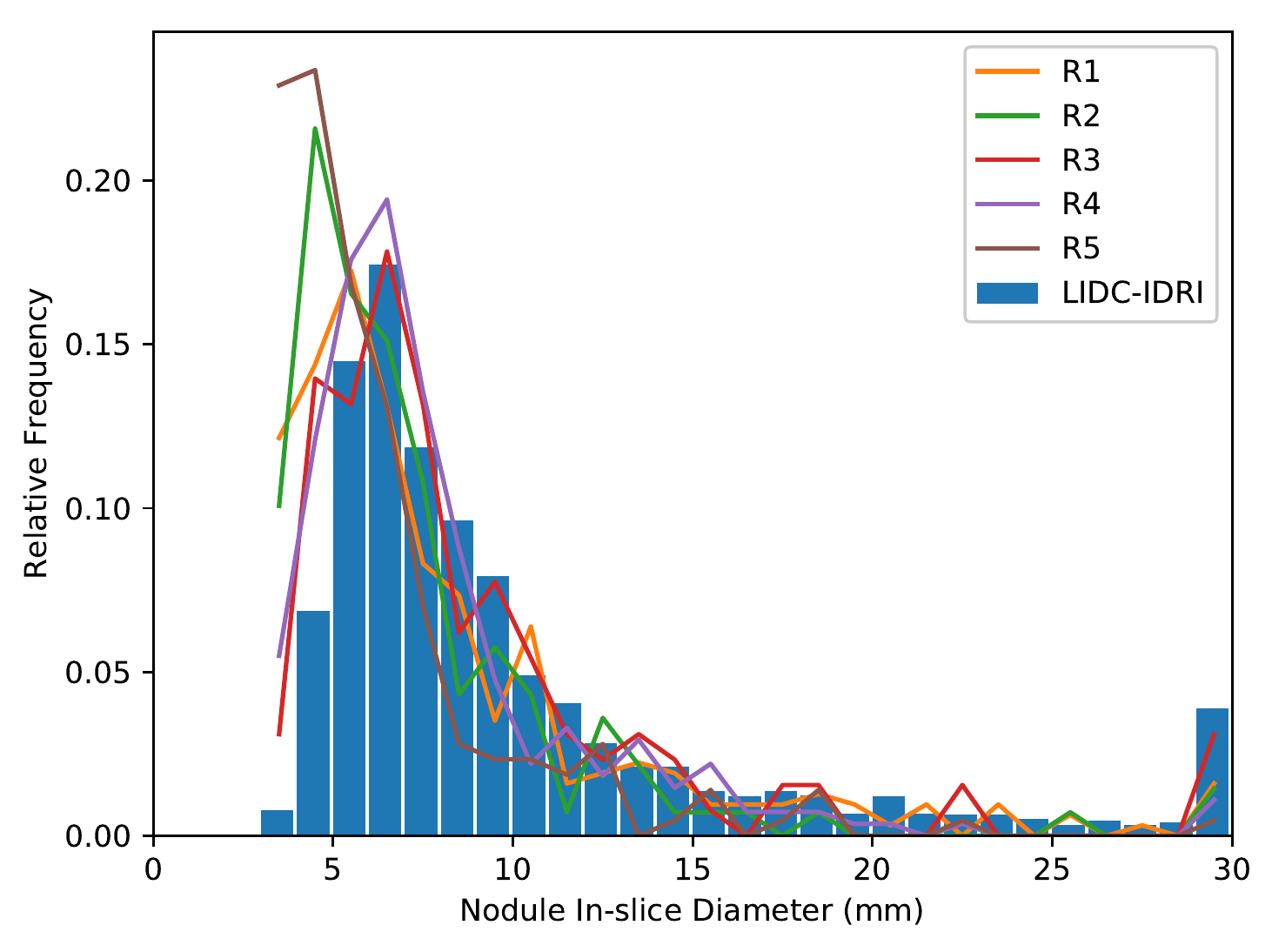}
		\vspace{-.6cm}
		\caption{}
		\vspace{-.2cm}
	\end{subfigure}
	\begin{subfigure}[c]{0.63\textwidth}
		\includegraphics[width=\textwidth,trim={0.2cm 0.3cm 0.3cm 0.3cm},clip]{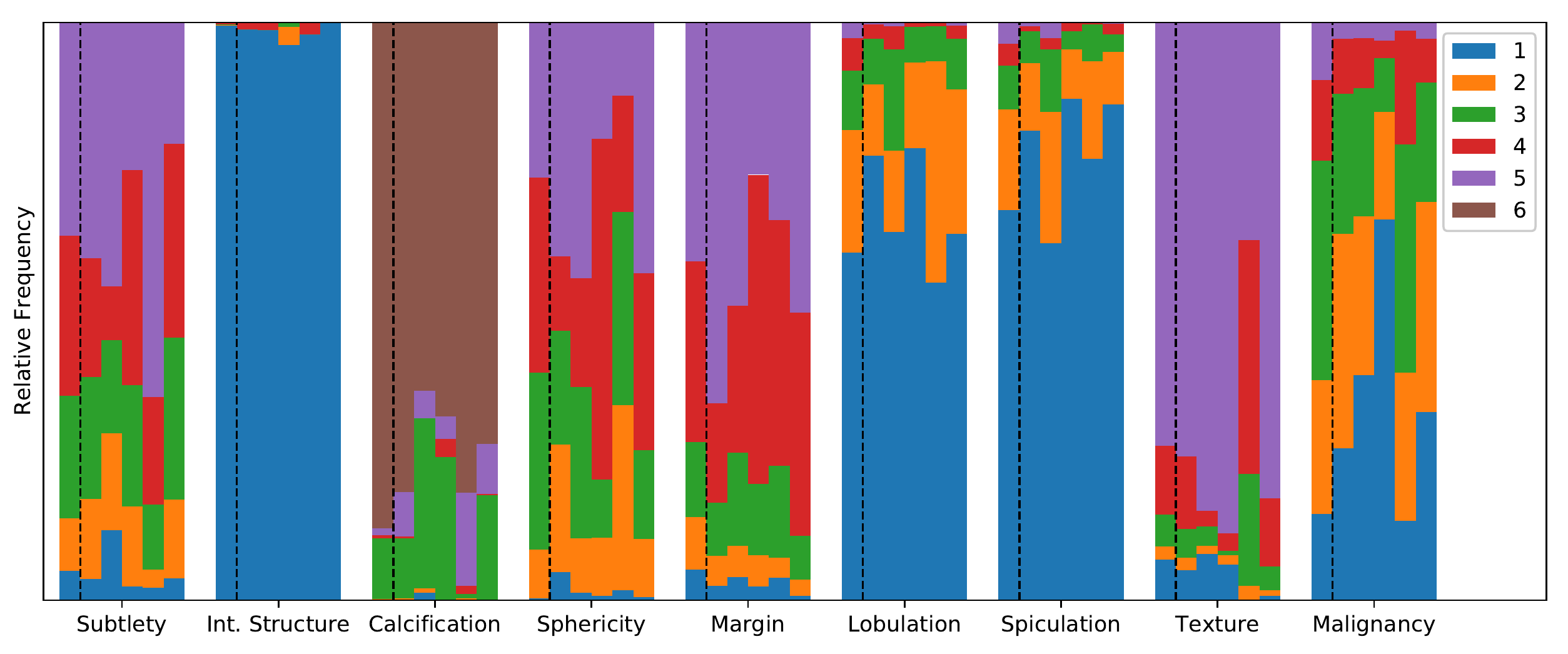}
		\caption{}
		\vspace{-.3cm}
	\end{subfigure}
	\caption{\small Nodule size and characterization distribution in LNDb and LIDC-IDRI. (a) Nodule in-slice diameter for LIDC-IDRI and R1 to R5. (b) Nodule characterization distribution for LIDC-IDRI (leftmost bar) and R1 to R5 (five rightmost bars). Colors correspond to each of the 1-6 ratings. Note that internal structure is rated in a 1-4 range, calcification in 1-6 and the other characteristics in 1-5.}
	\label{fig:LNDbSizeChar}
\end{figure*}

Figure \ref{fig:LNDbSizeChar} shows the nodule in-slice diameter and characteristics distribution in LNDb compared to LIDC-IDRI. In-slice diameter was determined as the largest distance across two points in any axial slice for nodules $\geq$3mm. It can be seen that the distribution in both size and characteristics follows that of LIDC-IDRI. However, more nodules $<$5mm have been annotated in LNDb, particularly by R2 and R5.

\subsection{Observer Variability}
Table \ref{tb:NodDtct_InterO} shows the nodule detection agreement $A_d$. It can be seen that the agreement is smaller than for LIDC-IDRI.

\begin{table}
	\centering
	\begin{tabular}{lc}
		\hline
		\vspace{-.2cm}&\\
		&$A_d$\\
		\hline
		\vspace{-.2cm}&\\
		R1 vs R2  &0.31\\
		R1 vs R3  &0.34\\
		R2 vs R3  &0.32\\
		R2 vs R4  &0.23\\
		R2 vs R5  &0.40\\
		R4 vs R5  &0.31\\
		All vs All&0.32\\
		\hline
		\vspace{-.2cm}&\\
		LIDC-IDRI&0.38\\
	\end{tabular}
	\\
	\centering
	Note that only pairs of radiologists with CTs in common are shown.
	\caption{Nodule detection agreement for LNDb and LIDC-IDRI.}
	\label{tb:NodDtct_InterO}
\end{table}

Table \ref{tb:NodSegm_InterO} shows the nodule segmentation agreement in terms of MAD, HD and Jaccard. It can be seen that the segmentation agreement is slightly higher for most radiologist pairs in comparison to LIDC-IDRI in terms of MAD and HD but lower for Jaccard. All metrics show a statistically significant difference when comparing all radiologists to LIDC-IDRI.

\begin{table}
	\centering
	\setlength\tabcolsep{.6em}
	\begin{tabular}{lccc}
		\hline
		\vspace{-.2cm}&&&\\
		&MAD (mm)&HD (mm)&Jaccard\\
		\hline
		\vspace{-.2cm}&&&\\
		R1 vs R2  &     0.46$\pm$0.26          &\hsig2.19$\pm$1.52$^\dagger$ &\hsig0.57$\pm$0.15$^\ddagger$\\
		R1 vs R3  &     0.49$\pm$0.19          &     2.45$\pm$1.65           &\hsig0.58$\pm$0.12$^\ddagger$\\
		R2 vs R3  &     0.53$\pm$0.18          &     2.26$\pm$1.44           &\hsig0.54$\pm$0.14$^\ddagger$\\
		R2 vs R4  &     0.32$\pm$0.18          &     1.51$\pm$0.39           &     0.58$\pm$0.17           \\
		R2 vs R5  &     0.28$\pm$0.04          &     1.58$\pm$0.52           &     0.62$\pm$0.07           \\
		R4 vs R5  &     0.41$\pm$0.19          &     1.88$\pm$0.89           &\hsig0.57$\pm$0.14$^\ddagger$\\
		All vs All&\hsig0.45$\pm$0.21$^\dagger$&\hsig2.13$\pm$1.35$^\ddagger$&\hsig0.57$\pm$0.14$^\ddagger$\\
		\hline
		\vspace{-.2cm}&&&\\
		LIDC-IDRI&0.48$\pm$0.38&2.92$\pm$3.15&0.66$\pm$0.13\\
	\end{tabular}
	\\
	\centering
	Note that only pairs of radiologists with CTs in common are shown.
	\caption{Nodule segmentation agreement (mean$\pm$standard deviation) for LNDb and LIDC-IDRI. Symbols $\dagger$ and $\ddagger$ indicate a statistically significant difference in comparison to LIDC-IDRI observer variability with p$<$0.05 and p$<$0.01, respectively.}
	\label{tb:NodSegm_InterO}
\end{table}

Table \ref{tb:NodChar_InterO} shows the nodule characterization agreement. It can be seen that, overall, the agreement is higher than for LIDC-IDRI except for calcification which has significantly lower agreement.

\begin{table*}
	\centering
	\setlength\tabcolsep{.6em}
	\begin{tabular}{lcccccccccccccccccc}
		\hline
		\vspace{-.2cm}&&&&&&&&&&&&&&&&&&\\
		&\multicolumn{2}{c}{Subtlety}&\multicolumn{2}{c}{Int. Structure}&\multicolumn{2}{c}{Calcification}&\multicolumn{2}{c}{Sphericity}&\multicolumn{2}{c}{Margin}&\multicolumn{2}{c}{Lobulation}&\multicolumn{2}{c}{Spiculation}&\multicolumn{2}{c}{Texture}&\multicolumn{2}{c}{Malignancy}\\
		&$A_c$&$\kappa_w$&$A_c$&$\kappa$&$A_c$&$\kappa$&$A_c$&$\kappa_w$&$A_c$&$\kappa_w$&$A_c$&$\kappa_w$&$A_c$&$\kappa_w$&$A_c$&$\kappa_w$&$A_c$&$\kappa_w$\\
		\hline
		\vspace{-.2cm}&&&&&&&&&&&&&&&&&&\\
		R1 vs R2&0.48&0.53&1.00&1.00&0.71&0.27&0.48&0.41&0.44&0.37 &0.56&0.28&0.57&0.33 &0.84&0.78&0.33&0.57\\
		R1 vs R3&0.53&0.58&0.95&0.00&0.86&0.66&0.35&0.33&0.33&0.31 &0.67&0.54&0.76&0.56 &0.83&0.60&0.45&0.57\\
		R2 vs R3&0.53&0.54&0.94&0.00&0.65&0.37&0.40&0.46&0.42&0.46 &0.52&0.25&0.65&0.60 &0.90&0.77&0.40&0.52\\
		R2 vs R4&0.57&0.48&0.86&0.00&1.00&1.00&0.43&0.31&0.00&-0.25&0.43&0.00&0.43&-0.10&0.14&0.00&0.14&-0.14\\
		R2 vs R5&0.33&0.41&0.92&0.00&1.00&1.00&0.42&0.41&0.25&-0.13&0.58&0.62&0.58&0.58 &0.67&0.00&0.33&0.30\\
		R4 vs R5&0.49&0.37&0.98&0.00&0.81&0.46&0.38&0.42&0.47&0.19 &0.63&0.46&0.68&0.32 &0.55&0.45&0.44&0.48\\
		All vs All&0.50&NA&0.97&NA&0.78&NA&0.40&NA&0.41&NA &0.60&NA&0.66&NA&0.72&NA&0.41&NA\\
		\hline
		\vspace{-.2cm}&&&&&&&&&&&&&&&&&&\\
		LIDC-IDRI&0.40&NA&0.99&NA&0.92&NA&0.35&NA&0.41&NA&0.49&NA&0.56&NA&0.69&NA&0.39&NA\\
	\end{tabular}
	\\
	\centering
	Note that only pairs of radiologists with CTs in common are shown.
	\caption{Nodule characterization agreement for LNDb and LIDC-IDRI. Note that in LIDC-IDRI only nodules $\geq$3mm were characterized and that for spiculation and lobulation the first 399 CTs contain incorrect labels and were thus excluded. NA: not applicable.}
	\label{tb:NodChar_InterO}
\end{table*}

Table \ref{tb:Fleischner_InterO} shows the scanwise agreement in terms of Fleischner score. Scanwise agreement is lower than for LIDC-IDRI, in spite of the fact that the agreement for Fleischner volume and texture classes is higher.

\begin{table}
	\centering
	\setlength\tabcolsep{.6em}
	\begin{tabular}{lccccccc}
		\hline
		\vspace{-.2cm}&&&&&&&\\
		&\multirow{2}{*}{$N_{CT}$}&\multicolumn{2}{c}{Follow-up}&\multicolumn{2}{c}{Volume}&\multicolumn{2}{c}{Texture}\\
		&&$A_c$&$\kappa_w$&$A_c$&$\kappa_w$&$A_c$&$\kappa_w$\\
		\hline
		\vspace{-.2cm}&&&&&&&\\
		R1 vs R2  &81 &0.58&0.42&0.80&0.83&0.94&0.74\\
		R1 vs R3  &83 &0.66&0.68&0.80&0.86&0.91&0.47\\
		R2 vs R3  &81 &0.65&0.57&0.74&0.81&0.95&0.74\\
		R2 vs R4  &11 &0.64&0.01&1.00&1.00&0.86&0.00\\
		R2 vs R5  &11 &0.82&0.40&1.00&1.00&0.92&0.00\\
		R4 vs R5  &157&0.68&0.57&0.85&0.87&0.90&0.62\\
		All vs All&235&0.65&NA  &0.82&NA  &0.92&NA  \\
		\hline
		\vspace{-.2cm}&&&&&&&\\
		LIDC-IDRI&1010&0.73&NA&0.81&NA&0.88&NA\\
	\end{tabular}
	\\
	\centering
	Note that only pairs of radiologists with CTs in common are shown.\\
	\caption{Scanwise agreement according to Fleischner guidelines for LNDb and LIDC-IDRI. $N_{CT}$ is the number of CTs analysed by each pair of radiologists. NA: not applicable.}
	\label{tb:Fleischner_InterO}
\end{table}

\subsection{Computer-Aided Annotation}
\begin{figure}
	\centering
	\includegraphics[width=\linewidth,trim={.2cm .3cm .3cm 0cm},clip]{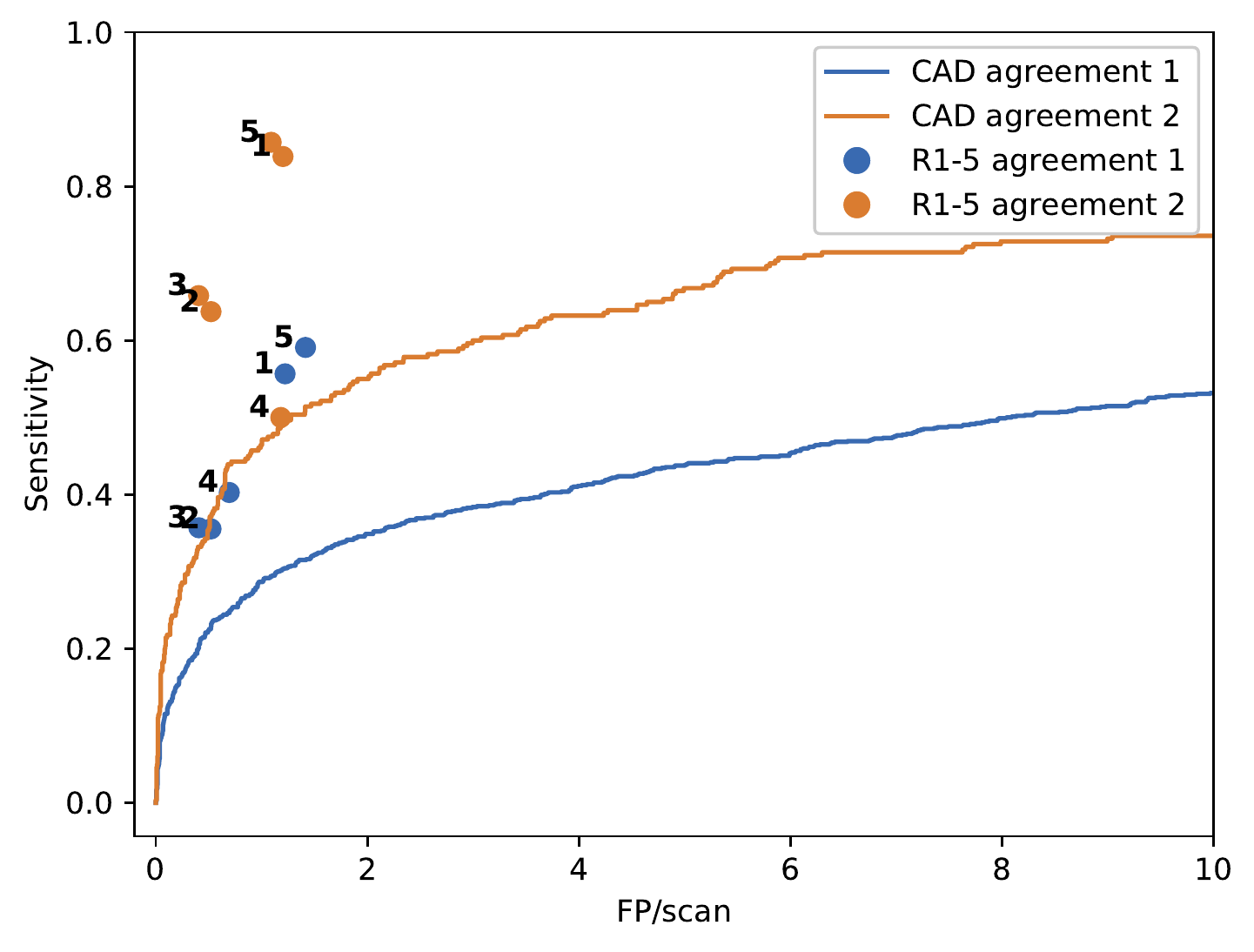}
	\caption{\small CAD and individual radiologist nodule detection performance for findings marked as a nodule with agreement level 1 and 2.}
	\label{fig:LNDb_CADDtct}
\end{figure}

\begin{figure*}
	\centering
	\includegraphics[width=\linewidth,trim={0cm 0cm 0cm 0cm},clip]{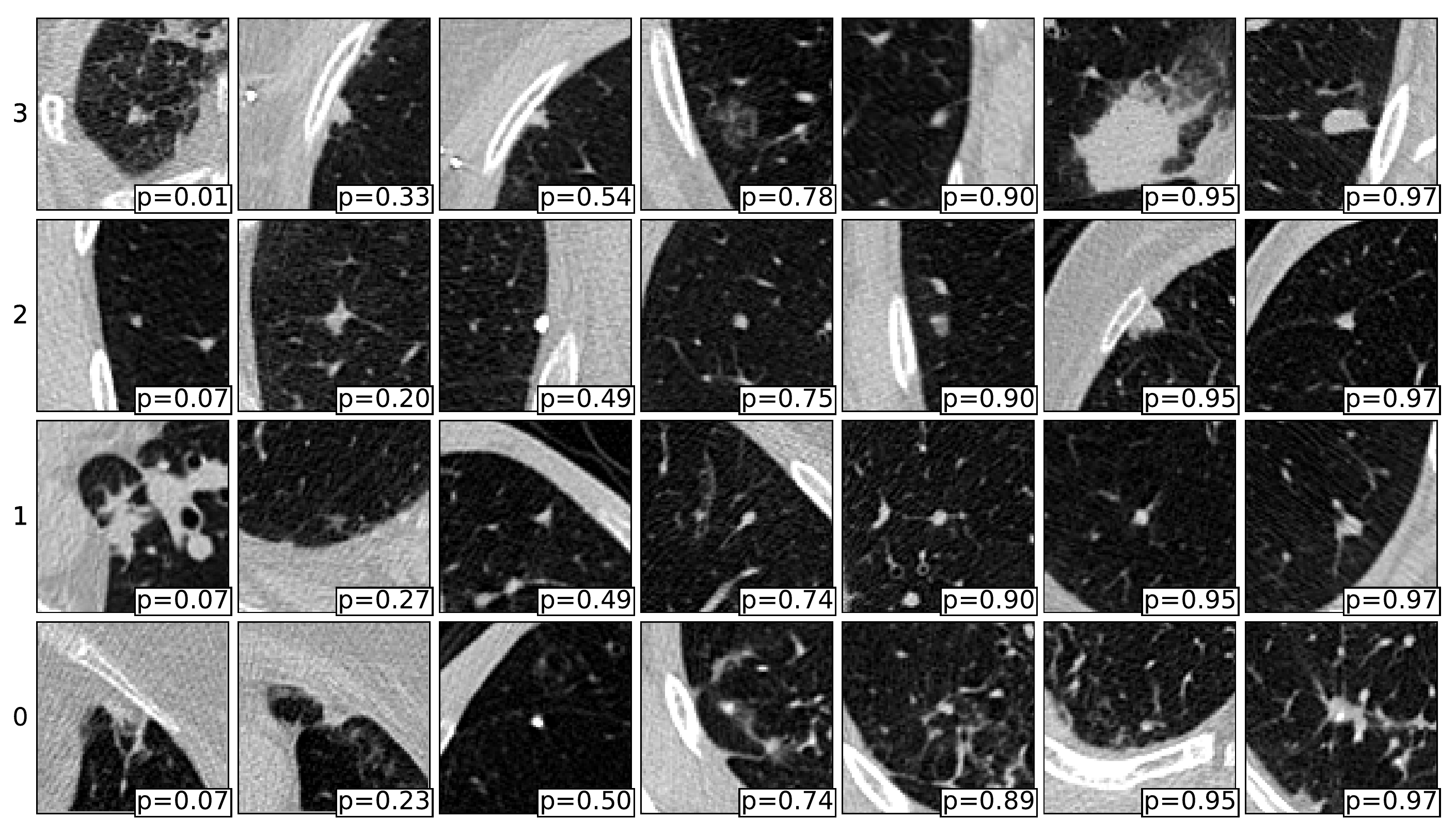}
	\caption{\small Central axial view (51$\times$51mm) of CAD detection examples on CTs annotated by 3 radiologists. Lines correspond to detection candidates with the same agreement level (findings annotated by 3, 2, 1 and 0 radiologists from top to bottom). Columns correspond to candidates with similar probability as given by the FP reduction algorithm (lower right corner of each frame). The seven probability levels correspond to a FP/scan level of 1/8, 1/4, 1/2, 1, 2, 4 and 8 for nodules with agreement level 2 (right to left).}
	\label{fig:LNDb_CADDtct_examples}
\end{figure*}

Figure \ref{fig:LNDb_CADDtct} shows the detection performance of CAD and each radiologist when considering the remaining radiologists as ground truth for each agreement level. Within the same agreement level, the average radiologist has a higher sensitivity than the CAD with 0.85 and 0.88 FPs per scan for agreement levels 1 and 2, respectively. The CAD system is able to obtain the same sensitivity as the average radiologist only at 5.99 and 5.80 FPs per scan for agreement level 1 and 2, respectively. Figure \ref{fig:LNDb_CADDtct_examples} shows examples of nodule candidates proposed by the automatic detection.

\begin{figure*}
	\centering
	\includegraphics[width=\linewidth,trim={0cm 0cm 0cm 0cm},clip]{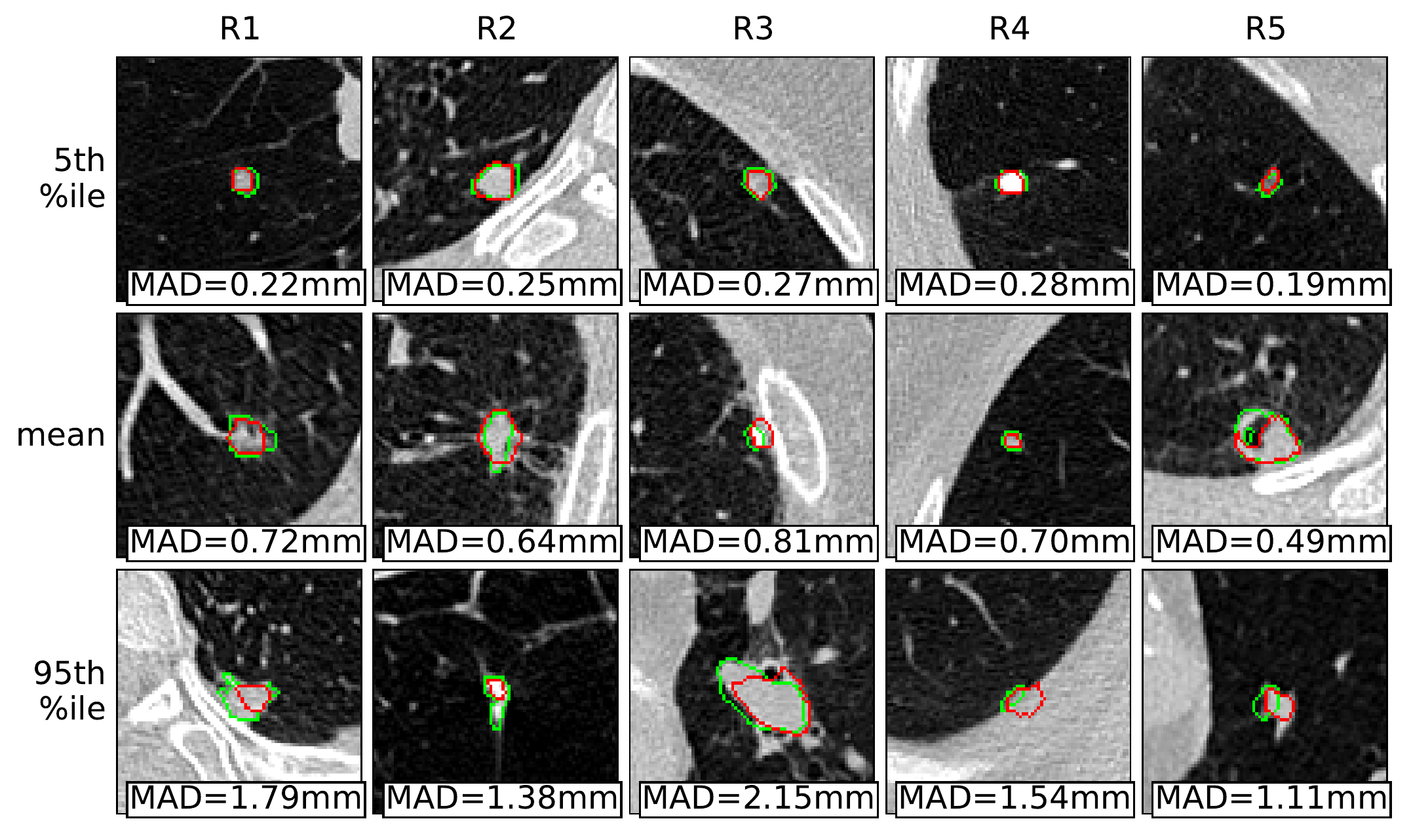}
	\caption{\small Central axial view (51$\times$51mm) of CAD segmentation examples (red) and ground truth annotations (green). Lines correspond to examples at the 5th percentile, mean and 95th percentile of MAD results obtained. Columns correspond to each radiologist. MAD obtained for each nodule shown at the lower right corner of each frame. Note that MAD is computed in 3D space whereas here only the central axial slice of each nodule is shown.}
	\label{fig:LNDb_CADSegm_examples}
\end{figure*}

Table \ref{tb:LNDb_CADSegm} shows the CAD segmentation performance when compared to the segmentations by each radiologist. Overall, there is a statistically significant difference in the agreement between the CAD and the radiologist annotations and the agreement observed among the radiologists in both the LIDC-IDRI and the LNDb databases. Figure \ref{fig:LNDb_CADSegm_examples} shows examples of nodule segmentations by each radiologist and the automatic segmentation.

\begin{table}
	\centering
	\setlength\tabcolsep{.6em}
	\begin{tabular}{lccc}
		\hline
		\vspace{-.2cm}&&&\\
		&\multicolumn{1}{c}{MAD (mm)}&HD (mm)&Jaccard\\
		\hline
		\vspace{-.2cm}&&&\\
		CAD vs R1 &0.72$\pm$0.66$^{\ddagger,\ast\!\ast}$&3.57$\pm$3.48$^{\ddagger,\ast\!\ast}$&0.51$\pm$0.18$^{\ddagger,\ast\!\ast}$\\
		CAD vs R2 &0.63$\pm$0.56$^{\ddagger,\ast\!\ast}$&2.94$\pm$2.78$^{\ast\!\ast}$         &0.50$\pm$0.17$^{\ddagger,\ast\!\ast}$\\
		CAD vs R3 &0.83$\pm$0.78$^{\ddagger,\ast\!\ast}$&3.80$\pm$4.19$^{\dagger,\ast\!\ast}$ &0.48$\pm$0.18$^{\ddagger,\ast\!\ast}$\\
		CAD vs R4 &0.70$\pm$0.44$^{\ddagger,\ast\!\ast}$&3.30$\pm$2.81$^{\ast\!\ast}$         &0.45$\pm$0.19$^{\ddagger,\ast\!\ast}$\\
		CAD vs R5 &0.49$\pm$0.36                   &2.46$\pm$2.27$^{\ast}$                    &0.55$\pm$0.17$^{\ddagger}$\\
		CAD vs All&0.67$\pm$0.57$^{\ddagger,\ast\!\ast}$&3.21$\pm$3.13$^{\dagger,\ast\!\ast}$ &0.50$\pm$0.18$^{\ddagger,\ast\!\ast}$\\
	\end{tabular}
	\caption{CAD nodule segmentation performance (mean$\pm$standard deviation). Symbols $\dagger$ ($\ast$) and $\ddagger$ ($\ast\ast$) indicate a statistically significant difference in comparison to LIDC-IDRI (LNDb) observer variability with p$<$0.05 and p$<$0.01 respectively.}
	\label{tb:LNDb_CADSegm}
\end{table}

\begin{figure*}
	\centering
	\includegraphics[width=\linewidth,trim={0cm 0cm 0cm 0cm},clip]{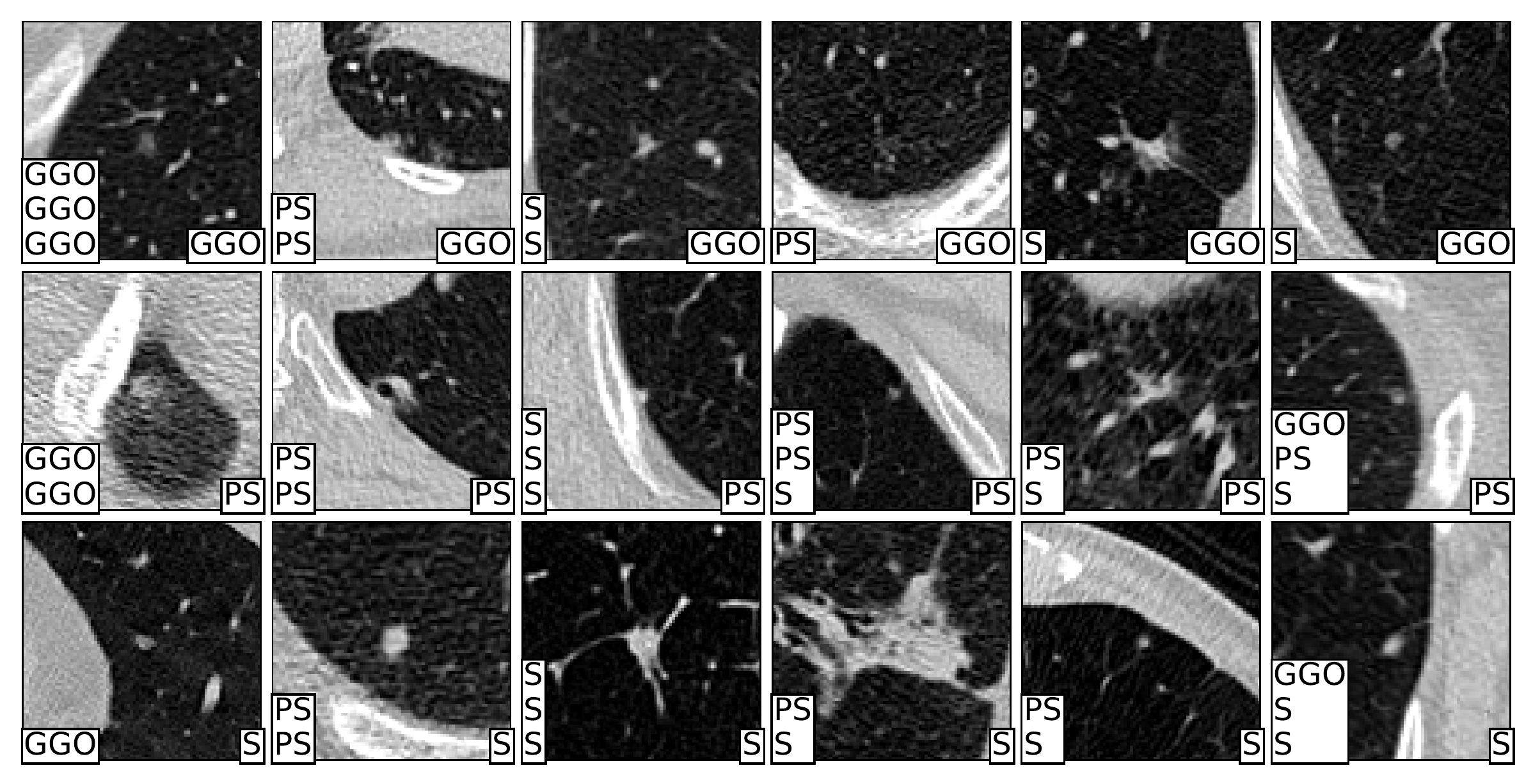}
	\caption{\small Central axial view (51$\times$51mm) of CAD texture characterization examples. Lines correspond to examples with equal texture class as given by the automatic algorithm. Different agreement levels in texture class by the annotating radiologists are illustrated. The texture class given by the automatic system is shown at the lower right corner of each frame whereas the texture classes given by the radiologists are shown at the lower left corner (lines correspond to multiple radiologists). S - Solid nodule, PS - Part-solid nodule, GGO - Ground glass opacity nodule.}
	\label{fig:LNDb_CADChar_examples}
\end{figure*}

Table \ref{tb:LNDb_CADFleischnerVolTex} shows the volume and texture characterization CAD performance according to Fleischner guidelines. The agreement in volume Fleischner classes is similar to the agreement between radiologists in LIDC-IDRI and LNDb. However, the agreement in Fleischner texture classes is significantly smaller than between radiologists. Figure \ref{fig:LNDb_CADChar_examples} shows examples of nodule texture characterization by the CAD compared to radiologist annotations.

\begin{table}
	\centering
	\setlength\tabcolsep{.6em}
	\begin{tabular}{lcccc}
		\hline
		\vspace{-.2cm}&&&&\\
		&\multicolumn{2}{c}{Volume}&\multicolumn{2}{c}{Texture}\\
		&$A_c$&$\kappa_w$&$A_c$&$\kappa_w$\\
		\hline
		\vspace{-.2cm}&&&&\\
		CAD vs R1 &0.77&0.72&0.79&0.61\\
		CAD vs R2 &0.80&0.74&0.82&0.67\\
		CAD vs R3 &0.69&0.65&0.76&0.52\\
		CAD vs R4 &0.77&0.74&0.76&0.54\\
		CAD vs R5 &0.90&0.84&0.74&0.15\\
		CAD vs All&0.80&0.75&0.77&0.51\\
	\end{tabular}
	\caption{CAD Fleischner volume and texture classification performance.}
	\label{tb:LNDb_CADFleischnerVolTex}
\end{table}

\begin{figure}
	\centering
	\includegraphics[width=\linewidth,trim={.3cm .3cm .3cm 0cm},clip]{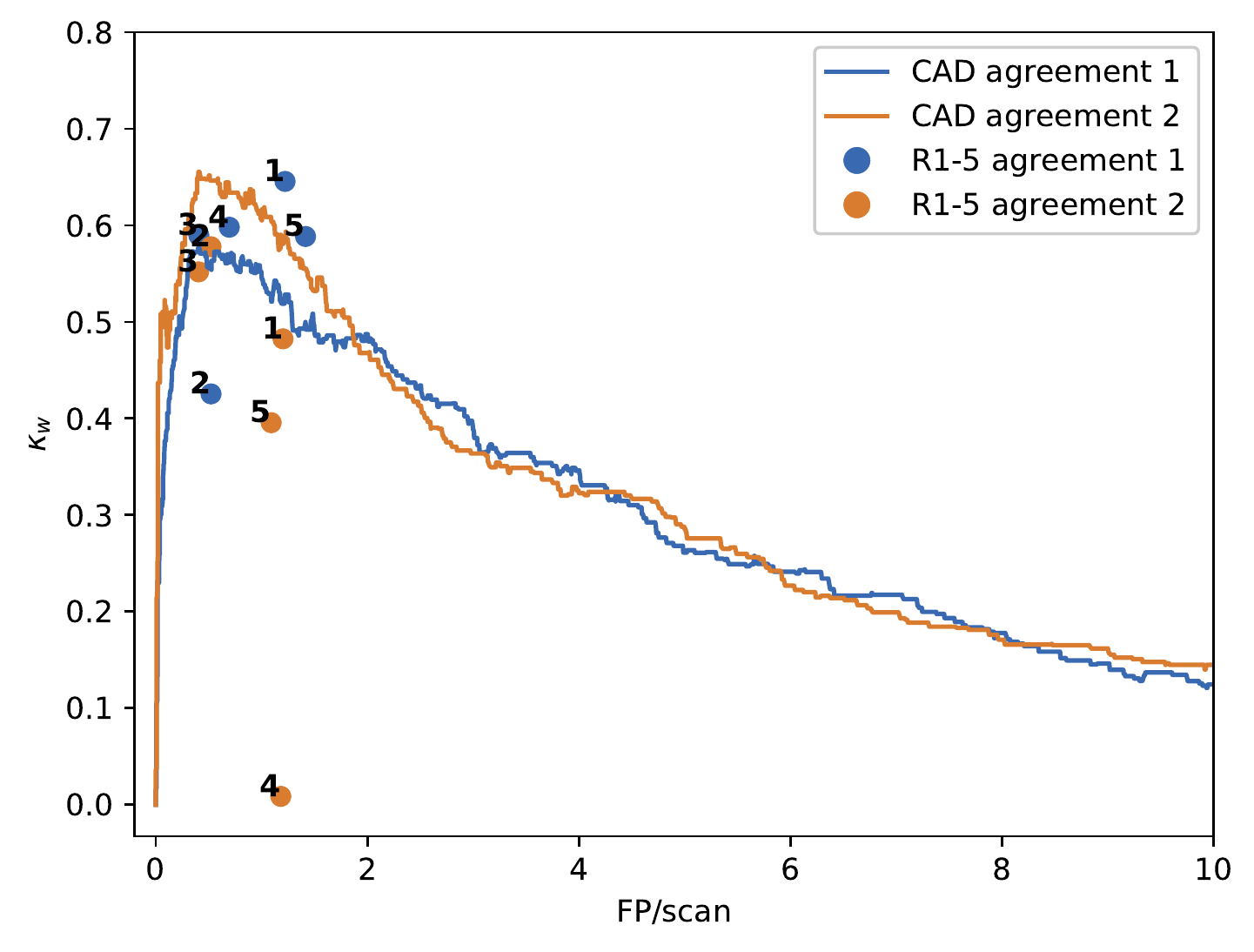}
	\caption{\small CAD and individual radiologist scanwise performance according to Fleischner guidelines considering findings marked as a nodule with agreement level 1 and 2.}
	\label{fig:LNDb_CADFleischner}
\end{figure}

Figure \ref{fig:LNDb_CADFleischner} shows the scanwise CAD performance according to Fleischner guidelines as well as the performance of each radiologist when considering the remaining radiologists as ground truth for each agreement level. Results are shown as function of FPs/scan according to the FP reduction threshold used. It can be seen that for both agreement levels the average radiologist agreement is similar to that of CAD within the same FP/scan level. Note that for R4 and R5 a limited number of CTs with agreement level 2 exist (11), which leads to the low $\kappa_w$ scores obtained.

\subsection{Collaborative Annotation Strategies}
\begin{figure*}
	\centering
	\begin{subfigure}[c]{0.475\textwidth}
		\includegraphics[width=\textwidth,trim={.2cm .3cm .3cm 0cm},clip]{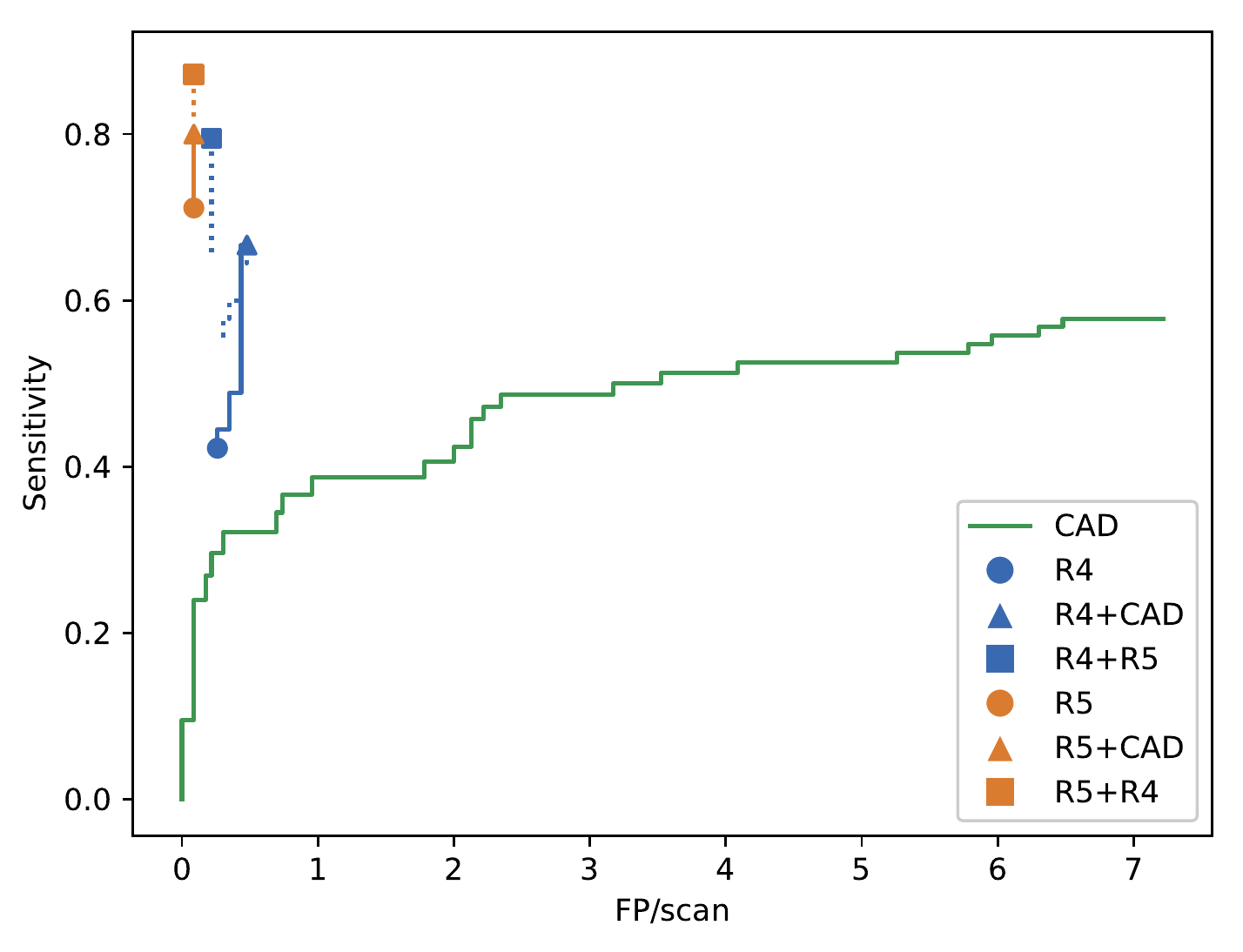}
		\vspace{-.6cm}
		\caption{}
		\vspace{-.2cm}
	\end{subfigure}
	\begin{subfigure}[c]{0.475\textwidth}
		\includegraphics[width=\textwidth,trim={.2cm .3cm .3cm 0cm},clip]{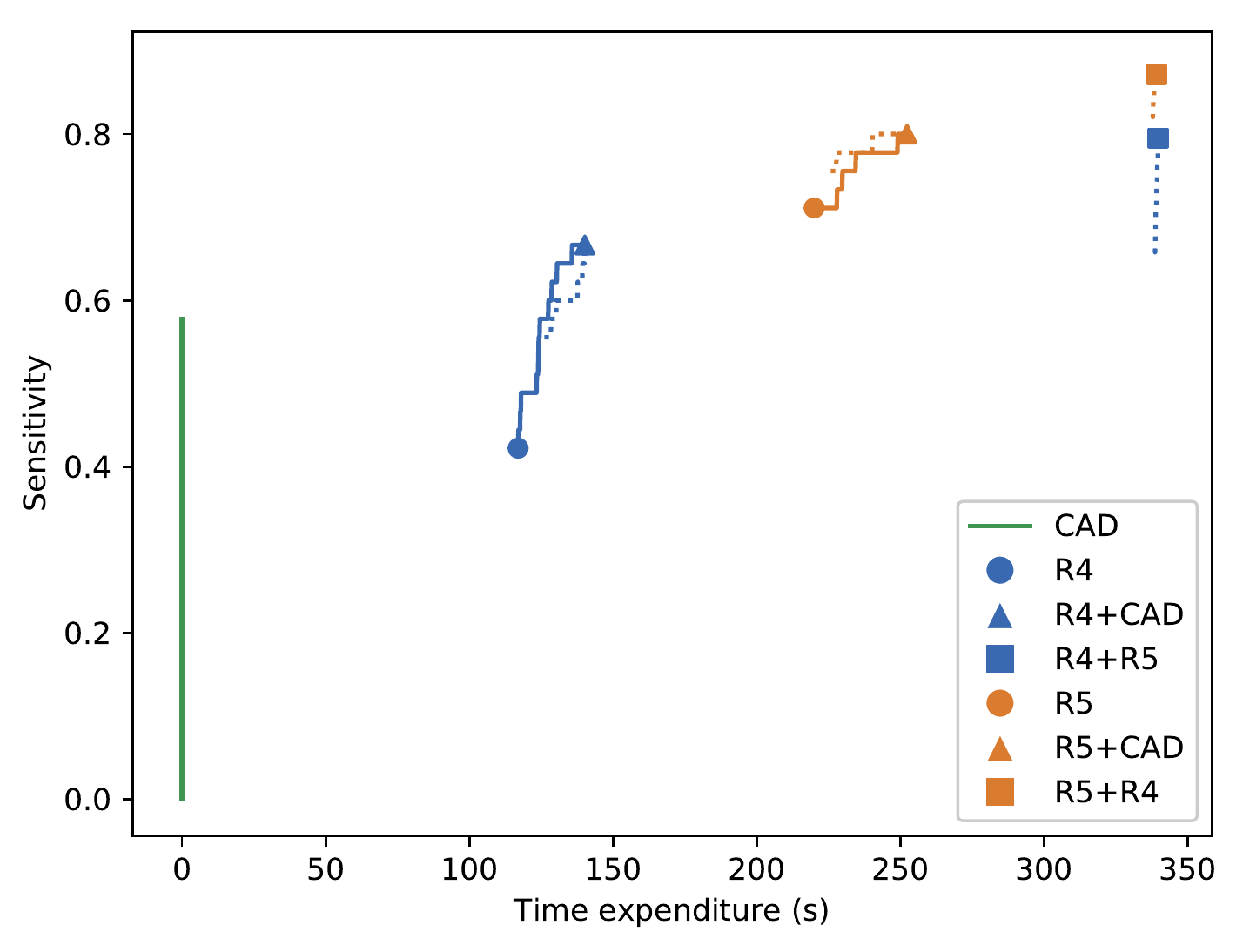}
		\vspace{-.6cm}
		\caption{}
		\vspace{-.2cm}
	\end{subfigure}
	\caption{\small Nodule detection performance for R4, R5, CAD and collaborative strategies. Full lines indicate performance at different FP reduction threshold levels and dotted lines indicate performance at different attention threshold levels.}
	\label{fig:LNDb_2ndDtct}
\end{figure*}

Figure \ref{fig:LNDb_2ndDtct} shows the detection performance of each annotation strategy. It can be seen that both radiologists have a superior performance to CAD but either of the collaborative approaches, using CAD or a second radiologist, give a significant boost to performance. In terms of time expenditure, the radiologist+CAD combination is the most efficient as it achieves high sensitivity with a small time investment in comparison to single radiologist annotation. Table \ref{tb:LNDb_2ndDtctFP} shows the anatomical structures identified as nodules by the CAD but as FPs by R4 and R5.

\begin{table}
	\centering
	\begin{tabular}{lc}
		\hline
		\vspace{-.2cm}&\\
		Vascular structures (arteries, veins)& 87 (52.4)\\
		Lymph nodes & 3 (1.8)\\
		Airway structures (bronchi, bronchioli, & \multirow{2}{*}{4 (2.4)}\\
		\hspace{.1cm}bronchial wall, bronchiectasis, etc.)&\\
		Parenchymal features &\\
		\hspace{.3cm} Atelectasis & 27 (16.3)\\
		\hspace{.3cm} Reticulation  (inter/intralobular septa)& 8 (4.8)\\
		\hspace{.3cm} Fibrosis & 2 (1.2)\\
		\hspace{.3cm} Ground glass opacities (nonspecific) & 5 (3.0)\\
		\hspace{.3cm} No visible/identifiable structure & 15 (9.0)\\
		Extrapulmonary structures&\\
		\hspace{.3cm} Bone & 10 (6.0)\\
		\hspace{.3cm} Other & 5 (3.0)\\
		\hline
		\vspace{-.2cm}&\\	
	\end{tabular}
	\caption{Anatomical structures identified as nodules by the CAD detection. Data are count (\%).}
	\label{tb:LNDb_2ndDtctFP}
\end{table}

Table \ref{tb:LNDb_2ndFleischnerVolTex} shows the volume and texture characterization performance according to Fleischner guidelines in comparison to ground truth for each of the annotation strategies considered. For both the texture and volume Fleischner classes it can be seen that either of the collaborative strategies increases the agreement when compared to single radiologist or CAD annotations.

\begin{table}
	\centering
	\setlength\tabcolsep{.6em}
	\begin{tabular}{lcccc}
		\hline
		\vspace{-.2cm}&&&&\\
		&\multicolumn{2}{c}{Volume}&\multicolumn{2}{c}{Texture}\\
		&$A_c$&$\kappa_w$&$A_c$&$\kappa_w$\\
		\hline
		\vspace{-.2cm}&&&&\\
		R4    &0.94&0.93&0.68&0.66\\
		R5    &0.90&0.75&0.94&0.64\\
		CAD   &0.98&0.93&0.71&0.55\\
		\hline
		\vspace{-.2cm}&&&&\\
		R4+CAD&1.00&1.00&0.79&0.76\\
		R4+R5 &0.97&0.94&0.84&0.83\\
		R5+CAD&0.92&0.75&0.95&0.91\\
		R5+R4 &0.92&0.75&0.95&0.91\\
	\end{tabular}
	\\
	\centering
	\caption{Volume and texture classification performance according to Fleischner guidelines for R4, R5, CAD and collaborative strategies.}
	\label{tb:LNDb_2ndFleischnerVolTex}
\end{table}

\begin{figure}
	\centering
	\includegraphics[width=.95\linewidth,trim={.3cm .3cm .3cm 0cm},clip]{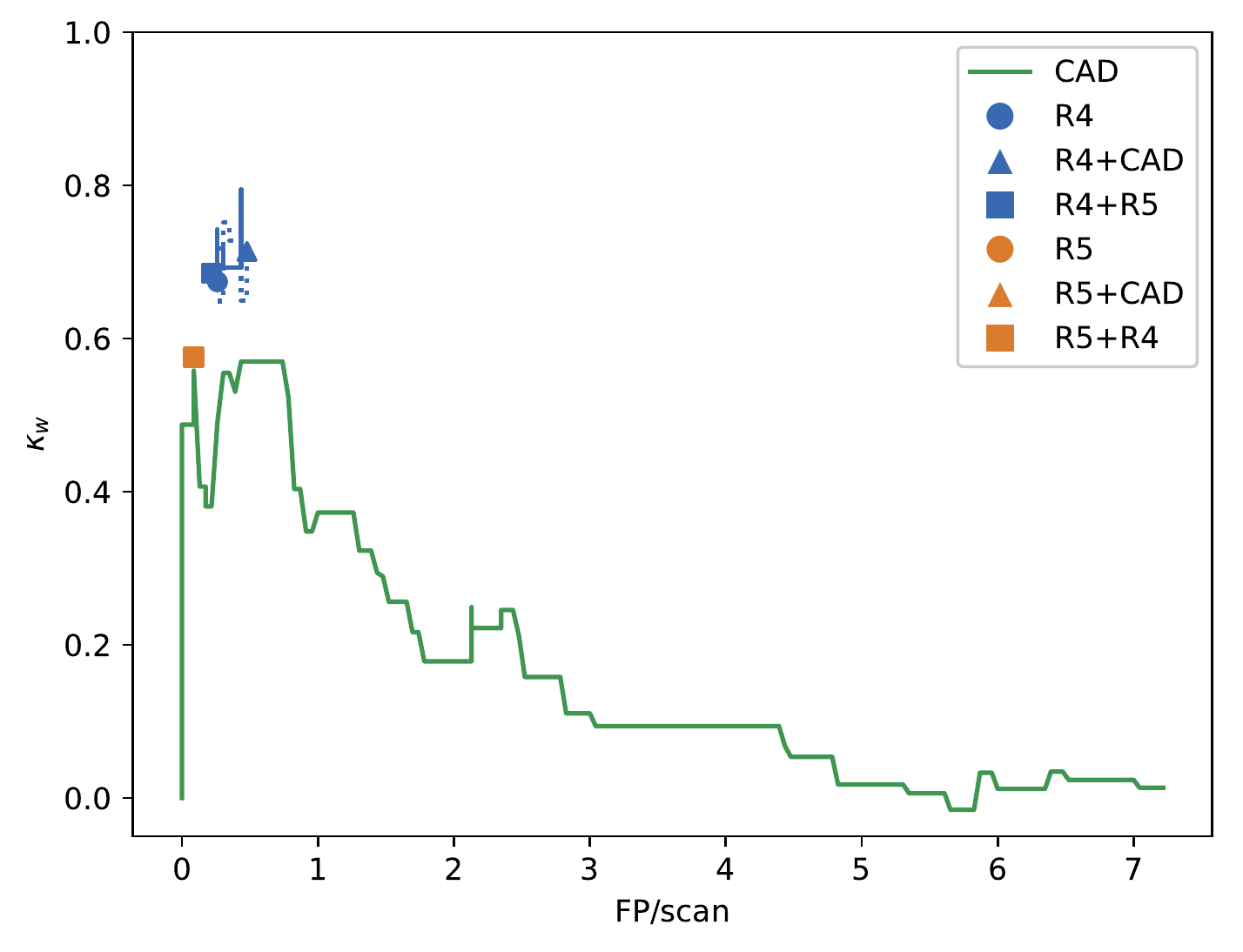}
	\caption{\small Scanwise performance for R4, R5, CAD and collaborative strategies. Full and dotted lines indicate performance at different FP reduction thresholds and different attention thresholds respectively.}
	\label{fig:LNDb_2ndFleischner}
\end{figure}

Figure \ref{fig:LNDb_2ndFleischner} shows the average scanwise performance of each annotation strategy according to Fleischner guidelines. It can be seen that, similarly to Figure \ref{fig:LNDb_CADFleischner}, manual annotation by a radiologist has a performance similar to CAD. However, collaborative strategies show only incremental improvement in performance for R4 and no improvement for R5.

\section{Discussion}
\subsection{LNDb Description}
In this study, the collection, annotation and analysis of a clinical CT database for nodule detection, segmentation and characterization is presented. While extensive public databases for this purpose exist, in particular the LIDC-IDRI database, the collection of local datasets and annotations can reveal surprising aspects in variability in the image collection and annotation processes and patient population with significant impact on the performance of CAD systems. This is especially true for deep learning methods, which depend to a large extent on the nature of the data used for training.

In the LNDb database, the variability in the annotation process was particularly emphasised as annotations were conducted solely in a single blinded manner, which replicates more closely the clinical reality where images are analysed by a single radiologist. Furthermore, in contrast to LIDC-IDRI, where radiologists were instructed to focus on nodules $\geq$3mm, the main task for radiologists in LNDb was to find all nodules, independent of size. This has led to a more diverse dataset, composed of a higher proportion of nodules $<$3mm than on LIDC-IDRI, as shown in Figure \ref{fig:LNDbSizeChar}. Accordingly, the characterization task was also performed for nodules $<$3mm, though this did not have an impact on the distribution of characteristics, as shown in Figure \ref{fig:LNDbSizeChar}. Some characteristics present extremely imbalanced distributions on LIDC-IDRI and this trend was repeated on LNDb in spite of the fact that radiologists were not trained in nodule characterization through examples from LIDC-IDRI.

\subsection{Observer Variability}
Overall, the single blind annotation protocol used to build LNDb has meant that observer variability was more accurately captured than in previous databases.

As shown in Table \ref{tb:NodDtct_InterO}, a smaller agreement in terms of nodule detection was obtained in comparison to LIDC-IDRI. This can be explained by the fact that LNDb annotations were obtained in a single blinded fashion whereas for LIDC-IDRI each radiologist would review the initial annotation after comparison to the annotations of other radiologists. As such, the LIDC-IDRI detection agreement can be solely attributed to decision error, i.e. deciding if each finding is a nodule or not. On LNDb, however, the detection agreement compounds the decision and fixation errors, i.e. the process of actually finding a nodule in the 3D CT image. Furthermore, the higher proportion of nodules $<$3mm can increase decision error, as the size of the finding can make decision more difficult \cite{gierada2008lung}.

In terms of nodule segmentation, Table \ref{tb:NodSegm_InterO} shows that the agreement in LNDb is higher than on LIDC-IDRI in MAD and HD. The fact that Jaccard is lower in LNDb is likely related to the higher proportion of smaller nodules, which can often have very small Jaccard. Nevertheless, while statistically significant, the difference is small in magnitude and has no impact on the volume Fleischner class agreement.

Nodule characterisation agreement (Table \ref{tb:NodChar_InterO}) and texture Fleischner class agreement is also larger on LNDb. The exception of calcification is probably related to the classes in this feature and its non-ordinal nature (1-Popcorn, 2-Laminated, 3-Solid, 4-Non-central, 5-Central, 6-Absent). In discussion with the radiologists involved in this study, it was clear that, to their understanding, calcification classes 3 and 5 were almost equivalent which can have led to the low agreement observed. This is corroborated in Figure \ref{fig:LNDbSizeChar}, where it can be seen that R2 and R3 mostly choose class 3 whereas R4 mostly chooses class 5 and R1 and R5 choose a mix of the two classes. As such, future studies should focus more on the presence/absence of calcification rather than on the type.

Overall, the fact that radiologists tend to agree more in the segmentation and characterization of nodules can be due to the fact that they belong to the same institution, have similar training and are using the same annotation tools, which was not the case in LIDC-IDRI.

For patient follow-up agreement, a lower agreement was obtained in LNDb, likely due to the lower nodule detection agreement observed. 

\subsection{Computer-Aided Annotation}
Figure \ref{fig:LNDb_CADDtct} shows the nodule detection performance of each radiologist and the CAD system when considering the remaining radiologists as ground truth for agreement levels 1 and 2. The degree of variability observed in LNDb is clearly expressed given the low sensitivity observed for radiologists when compared to previous studies \cite{setio2017validation}. Furthermore, it can be seen that the CAD has a comparatively poor performance, only achieving the average radiologist sensitivity at a relatively high FP/scan rate. While the high observer variability can have played a role, a performance closer to the average radiologist was expected, especially considering that an identical network was able to obtain a sensitivity of 0.926 at 0.25FP/scan on a subset of LIDC-IDRI \cite{aresta2018towards}. The main reasons for these results are probably related to the fact that the nodule detection network was trained uniquely on LIDC-IDRI. Firstly, only nodules $\geq$3mm were considered for training. The higher prevalence of smaller nodules on LNDb could thus have played a significant role in the decreased perfomance, given that the network was not trained particularly for these nodules. Secondly, and perhaps most importantly, while LIDC-IDRI is extensive, significant differences in the image acquisition or population characteristics can have led to a decreased performance. The slice thickness distribution, for example, is significantly different between the two databases, which may lead to detrimental performance. This highlights the need for fine tuning of deep learning methods before their application to specific cases and populations, showing that simply increasing the size of the dataset is not always the path towards increasing performance and that fine tuning and the setting of the problem must be taken into account.

Table \ref{tb:LNDb_2ndDtctFP} gives an insight into what anatomical features are wrongly identified as nodules by the CAD. While the amount of vascular FPs is not surprising given the 2D similarity of solid nodules and vessels, it was expected that the 2.5D and 3D nature of the detection and FP reduction networks would contribute to a smaller proportion of vascular FPs. Future approaches should thus take this shortcoming into account and try to incorporate further 3D information and/or specific architecture to target vascular FPs.

In terms of nodule segmentation, a statistically significant difference between the CAD performance and the observer variability on the LNDb and LIDC-IDRI was found. However, the difference is not significantly detrimental for Fleischner volume classes as the average CAD performance of 0.80 is similar to the agreement between radiologists on both LNDb and LIDC-IDRI (0.82 and 0.81 respectively). Furthermore, in contrast to the behaviour observed for nodule detection, the segmentation network's application to a different database did not have a detrimental effect as a Jaccard index of 0.48$\pm$0.19 was reported for the LIDC-IDRI database in \cite{aresta2019iw}.

A performance in terms of $A_c$ of 0.77 was obtained for texture characterization, which is inferior to the average observer agreement in LNDb and LIDC-IDRI. Nevertheless, as for nodule segmentation, this performance is similar to the reported performance for LIDC-IDRI (0.751$\pm$0.035) \cite{ferreira2018convolutional}.

In terms of scanwise performance, Figure \ref{fig:LNDb_CADFleischner} shows that there seems to be no significant difference between the CAD and the average radiologist. While this is in stark contradiction with the low performance of the CAD detection performance, one must take into account the fact that only the most suspicious nodules have an implication in terms of Fleischner score. As such, considering for example a patient with several nodules among which one is particularly large, it is sufficient for the CAD system to correctly detect and classify that nodule to obtain a correct Fleischner score. The fact that the FP/scan level with highest scanwise performance is relatively low shows exactly this fact. A similar detection sensitivity to the radiologist is not crucial, as long as the most suspicious nodules are correctly detected with a low number of FPs.

Taking into account the performance obtained at each stage and framing it in the greater picture of Fleischner follow-up guidelines thus gives a clearer picture of the role CAD systems may have in CT lung cancer screening. It is shown that state-of-the-art CAD systems are able to have a performance comparable to the average radiologist. This is in spite of the fact that the nodule detection performance in particular was rather poor in this study. While it might be tempting to conclude that techniques such as fine tuning could lead to an improved CAD scanwise performance due to improved nodule detection, this might not be the case. In fact, the superior detection performance observed for the radiologists did not lead to a superior performance in follow-up. As such, and even though improvement of the overall nodule detection performance is crucial to improve radiologists' trust in CAD systems, instead of focusing on marginal gains in detection performance, the community should focus on the accurate detection and characterization of those nodules known to be more associated with malignancy, namely solid and part solid $\geq$100mm\textsuperscript{3} nodules. Nevertheless, detection of smaller nodules could come to play a more important role in follow-up scenarios, as the change in size and characteristics of a nodule are a strong indicator of malignancy.

\subsection{Collaborative Annotation Strategies}
Figure \ref{fig:LNDb_2ndDtct} shows the results obtained for nodule detection in individual and collaborative strategies for a subset of LNDb. As in Figure \ref{fig:LNDb_CADDtct}, it is shown that the CAD has a lower performance than the average radiologist. Furthermore, there is a significant difference between R4 and R5 in terms of sensitivity, with R5 identifying a larger proportion of nodules. Any of the collaborative strategies significantly improve performance, especially for R4, with double radiologist strategies being the most successful. However, given the poor performance of the CAD system in this particular setting, this was to be expected.

In terms of the time spent analysing the image, it can be seen that R5 takes approximately double the amount of time as R4, which justifies the increased sensitivity, as R5 analyses the image more carefully. When comparing the two collaborative approaches, it can be seen that having a second opinion from CAD can significantly boost performance without having a large impact on the overall time spent. A double radiologist strategy has obviously a greater penalty on time spent, with almost 6 minutes per CT, in comparison to over 2 minutes and over 4 minutes spent by R4 and R5 respectively when receiving suggestions from CAD. Interestingly, neither the FP reduction nor the eyetracking data seem to be overwhelmingly beneficial in excluding nodules that do not require revision by a radiologist. Nevertheless, an approach where radiologists only revise nodule candidates from CAD which had not been observed before (eyetracking time of 0s) seems of particular interest with a sensitivity improvement of 0.13 and 0.04 for R4 and R5, respectively, through the revision of 3.0 and 1.3 CAD findings per scan respectively.

For segmentation and characterization, analysed in terms of the Fleischner classes, Table \ref{tb:LNDb_2ndFleischnerVolTex} shows that for both volume and texture there is an improvement in accuracy in either of the collaborative strategies. However, for nodule volume, improvements are quite small in magnitude given the already high accuracy of both radiologists and CAD.

Looking at the scanwise classification in terms of Fleischner guidelines for follow-up, it can be seen that the CAD has a performance similar to individual radiologists. Interestingly, even though R4 has a much smaller sensitivity for nodule detection, it has a slightly higher performance on follow-up, once more highlighting the importance of finding the `right' nodules, rather than finding all nodules. In regard to the collaborative strategies, while there are marginal performance gains for R4, these might be due to the low sample size of this experiment.
\subsection{Limitations}
While the results of this study are promising, there are important limitations to be considered. First, regarding LNDb, though one of its strenghts is the fact that it represents the clinical reality of a particular time and place, this is also a limitation as the results obtained might not be reproducible in other radiology departments. Furthermore, because the data was annotated in a single blind manner, there is no absolute ground truth. While this could be improved through a revision of all annotations by additional radiologists, this did not fall within reasonable effort for the scope of this study. Of course, this has a strong impact on the results and their interpretation.

Secondly, regarding the CAD systems used, the aim of this study was not to conduct an extensive review of state-of-the-art systems in literature. The algorithms used were chosen as they were deemed representative of the overall trends in literature. While other methodologies could have given origin to different conclusions, this was outside the scope of this study.

Third, regarding the collaborative annotations strategies studied, the low number of CTs used in this experiment limited the conclusions that could be drawn. Furthermore, while in this study two obvious collaborative strategies considering the tools available were tested, other more complex strategies could be designed which could be more successful. While only the eyetracking time per region was considered in this study, there is additional data that could be extracted such as the gaze patterns and fixation lengths that could provide further information. Furthermore, and given the objective of the Fleischner guidelines, one could take this into account by suggesting nodules that would change the Fleischner class if considered to be a nodule by the revising radiologist.

\section{Conclusion}
In conclusion, this study presents a novel database for research on several aspects of CT lung cancer screening: nodule detection, segmentation and characterization. Furthermore, the recording of the gaze patterns of radiologists when reading the images could hold important information useful for CAD and collaborative strategies.

By applying state-of-the-art detection, segmentation and characterization methods, it was shown that current CAD systems can classify a patient according to Fleischner follow-up guidelines as accurately as radiologists. Nevertheless, the training of deep learning methodologies for nodule detection can play a crucial role in performance and adaptation to the local characteristics in population, image acquisition, etc. is extremely important. Furthermore, within the three tasks, nodule detection was identified as the current biggest challenge, and thus the task where the biggest improvements can be made to obtain a better follow-up performance.

Finally, different collaborative strategies were tested in a subset of the data, showing that current CAD methodologies, even without fine tuning to a local reality, can be a valuable tool to increase sensitivity in nodule detection, without significantly increasing the burden for clinicians, especially if the collaboration between the two can be adequately designed. Nevertheless, this was not verified in Fleischner follow-up classification, where collaborative strategies did not lead to a significant improvement in comparison to individual radiologist annotation or state-of-the-art CAD systems.

\section*{Acknowledgment}
This work was financed by the European Regional Development Fund (ERDF) through the Operational Programme for Competitiveness - COMPETE 2020 Programme and by National Funds through the Portuguese Funding agency, FCT - Funda\c{c}\~ao para a Ci\^encia e Tecnologia within project PTDC/EEI-SII/6599/2014 (POCI-01-0145-FEDER-016673) and by the FCT grant contract SFRH/BD/120435/2016.

\ifCLASSOPTIONcaptionsoff
  \newpage
\fi

\bibliographystyle{IEEEtran}
\bibliography{IEEEabrv,bibliography}

\end{document}